\documentclass[journal=jctcce,manuscript=article]{achemso}
\usepackage[T1]{fontenc}
\usepackage[utf8]{inputenc}
\usepackage{graphicx}
\usepackage{amsmath,amssymb}
\usepackage{bm}
\usepackage{booktabs}
\usepackage{url}
\usepackage{nicefrac}
\usepackage[version=4]{mhchem}
\setkeys{acs}{articletitle=true}

\author{Evelio Francisco}
\email{evelio@uniovi.es}
\author{Jos\'e Antonio Qui\~nonero}
\author{\'Angel Mart\'{\i}n Pend\'as}
\email{ampendas@uniovi.es}
\affiliation[Universidad de Oviedo]{Departamento de Qu\'{\i}mica F\'{\i}sica y
Anal\'{\i}tica, Facultad de Qu\'{\i}mica, Universidad de Oviedo, 33006 Oviedo, Spain}

\title{Quadrature-Free Radial Intracule Densities: A Bessel-Harmonic Formulation for Gaussian Basis Sets}

\begin{document}

\begin{tocentry}
\includegraphics[width=3.25in,height=1.75in]{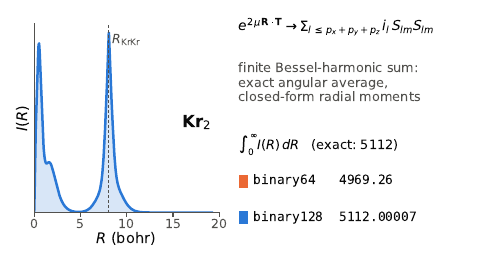}
\end{tocentry}

\begin{abstract}
The radial intracule density $I(R)$, the probability distribution of the
interelectronic distance, is a compact and physically transparent picture of
electron correlation, but its evaluation for molecules has always relied on a
numerical angular quadrature over the orientation of the interelectronic vector.
We show that this quadrature can be removed altogether for Cartesian Gaussian
basis sets and arbitrary second-order density matrices. Starting from the
Hermite Gaussian expansion of the products of basis functions, the coupling
exponential $e^{2\mu\bm{R}\cdot\bm{T}}$ that links radial and angular variables
is expanded in modified spherical Bessel functions and real spherical harmonics,
and the expansion terminates rigorously at $l=p_x+p_y+p_z$, where the $p$'s are the Cartesian
powers arising from the Hermite expansion. The angular average
is then exact, the atomic limit reduces to a finite sum of terms
$C_iR^{n_i}e^{-\mu R^2}$, and the radial moments $\int_0^\infty R^nI(R)\,dR$
are given in closed form through the confluent hypergeometric function
${}_1F_1$. The scheme is exact in principle and fails only arithmetically:
we identify the catastrophic cancellations in the angular-momentum coupling
that ruin binary64 evaluations for heavy atoms and remove them by elevating the
cancelling chain to binary128. Sum rules for the total, Coulomb and
exchange-correlation intracules are recovered to seven to eight significant
digits for systems from LiH to \ce{Kr2}, whereas Euler--Maclaurin and Lebedev
quadratures err by up to 22\% for stretched or multicenter molecules.
\end{abstract}

\section{Introduction}
\label{sec:introduction}
The concept of the intracule was introduced by Coulson and Neilson in the
early 1960s as a means of quantifying the effect of electron correlation on the
relative motion of two electrons \cite{Coulson1961}. In that seminal work, they
defined the Coulomb hole as the difference between the distribution function of
the interelectronic distance computed from the exact wave function, or the best
available approximation to it, and the corresponding distribution obtained from
the Hartree-Fock wave function, and they evaluated its radius, depth, and volume
for the ground state of helium. In its modern form, the
intracule probability density $I(\bm{R})$ is defined as the
probability density associated with the interelectronic distance vector
$\bm{R} = \bm{r}_2 - \bm{r}_1$, and it is obtained by integrating the diagonal
second-order density matrix $\rho^2(\bm{r}_1,\bm{r}_2)$ over all spatial and
spin coordinates except the relative coordinate $\bm{R}$.
Together with its momentum-space counterpart and
with the extracule density $E(\bm{R})$ \cite{Koga1998a,Koga1998b},
which describes the distribution of the
center of mass of an electron pair, the intracule provides a representation of
electron correlation that is at once global and intuitively appealing. The
coordinates that underlie both densities were introduced into density matrix
theory by Coleman \cite{Coleman1967}, and the formal properties of the
intracule, including its normalization, its moments, and the electron-electron
cusp condition that it must satisfy at the origin, were established by Thakkar
and Smith in a series of papers
\cite{ThakkarSmith1976,ThakkarSmith1977,Thakkar1984a,Thakkar1984b} and
reviewed by Thakkar in the late 1980s \cite{Thakkar1987}.

The intracule has since been applied to a remarkably broad range of problems,
and its usefulness goes well beyond a purely descriptive role. Fermi and
Coulomb holes have been examined in atoms \cite{Boyd1973,Boyd1974,Smith1971}
and used to analyze the angular aspects of electron correlation
\cite{Boyd1982}, while the parallel-spin counterpart of the Coulomb
hole (the Fermi hole) has been studied by Boyd and Coulson \cite{Boyd1974} and
by Ugalde and Boyd \cite{Ugalde1985}. Because $I(\bm{R})$ encodes the
probability of finding two electrons at a given separation, it provides a
physically transparent picture of the bonding process
\cite{RodriguezMayorga2018a} and reveals the topological fingerprints of the
Fermi and Coulomb holes in polyatomic molecules \cite{Boyd1988,Sarasola1990}.
The basis-set dependence of the intracule has
been analyzed by Dom\'inguez and co-workers in CO$_2$ \cite{Dominguez1992},
and the intracule has been used
to construct entirely new
theoretical frameworks for the correlation energy \cite{Gill2003,Bernard2008}.
On the methodological side, a wide range
of foundational analytical formulations of the intracule density for
polyatomic molecules has been developed over the years
\cite{Ugalde1991,Ugalde1985,Boyd1988,Sarasola1990} and later
highly optimized numerically \cite{ONeill2003}.
Finally, the intracule has proven essential for analyzing dynamical versus
nondynamical correlation \cite{ViaNadal2019,RamosCordoba2016,RamosCordoba2017},
for assessing the performance of approximate exchange-correlation functionals,
and for exploring the role of relativistic effects \cite{RodriguezMayorga2022},
as well as for understanding bonding
\cite{RodriguezMayorga2018a} and the topology of the Fermi and Coulomb
holes \cite{Boyd1988,Sarasola1990,ViaNadal2022}.

In atomic systems, the topographic analysis
of the intracule and extracule densities and of their Laplacian fields has
provided a two-electron perspective on the shell structure of atoms. Sarasola,
Dom\'{\i}nguez, Aguado, and Ugalde \cite{Sarasola1992} showed that the odd-numbered
zeros of $\nabla^2 E(\bm{R})$ and the radii corresponding to minima in
$\nabla^2 I(\bm{R})$ reproduce the shell structure predicted by the Bohr theory,
and that the latter match closely the corresponding minima in the Laplacian of
the one-electron density $\nabla^2 \rho(\bm{r})$ reported by Shi and Boyd
\cite{ShiBoyd1988} and by Sagar, Ku, Smith, and Simas \cite{Sagar1988}.
These results established that genuine two-electron
densities contain at least the same information as the one-electron density,
and that the intracule and extracule spaces are natural settings for the
analysis of electron-pair structure in atoms.

In molecular systems, the lack of spherical symmetry introduces an anisotropy
that must be taken into account. Sharma and Thakkar \cite{Sharma1984b} studied
the anisotropic Coulomb hole in H$_2$ and introduced the longitudinal and
transverse projections of the anisotropic Coulomb hole, showing that in H$_2$
Coulomb correlation has a greater effect on the longitudinal intracule than on
the transverse one. Boyd, Sarasola, and Ugalde \cite{Boyd1988} computed the
spherically averaged intracule density of H$_2$ with minimal and split-valence
basis sets and analyzed its dependence on the internuclear separation. They
showed that the restricted Hartree-Fock intracule density exhibits an
incorrect ridge at small interelectronic distances when the internuclear
separation goes to infinity, due to the incorrect dissociation of H$_2$, and
that configuration interaction calculations restore the correct qualitative
form. Wang, Tripathi, and Smith \cite{Wang1992} examined the influence of
electron correlation on the anisotropic intracule and extracule densities of
Ne, H$_2$, and LiH, using SCF and SDCI wave functions, and showed that electron
correlation makes the pair distributions more diffuse and corrects the
overestimation of the ionic structure that is characteristic of the
Hartree-Fock approximation. A complementary decomposition was proposed by
Fradera, Dur\'an, and Mestres \cite{Fradera2000}, who separated the electron-pair
density into uncorrelated and correlated components and analyzed the
corresponding components of the intracule, the extracule, and their Laplacians.
Their work showed that Fermi correlation accounts for the main topological
features of the correlated components, while Coulomb correlation increases the
locality of the electron-pair density.

The computational challenge posed by the evaluation of intracule and extracule
densities on large grids of points was addressed by Cioslowski and Liu in a
well-known contribution to this field \cite{Cioslowski1996},
in which they proposed a fast algorithm
based on precomputed intermediates. Their approach shifts a large portion of
the numerical effort from the grid-dependent to the grid-invariant phase of the
calculation, achieving speed-ups of 50 to 100 times over conventional
algorithms through the use of efficient two-stage integral screening. Cioslowski
and Liu also showed that, for regular grids with shared components of the
Cartesian coordinates, the primitive quartet contributions factorize in such a
way that the computational cost becomes proportional to the cubic root of the
number of grid points. Their work constitutes an important antecedent to the
present formulation, since it demonstrates that a judicious organization of the
calculation can drastically reduce the cost of evaluating these densities,
although it retains a numerical angular integration over the orientation of
$\bm{R}$.

Lee and Gill \cite{Lee1999} developed an alternative formulation within the \texttt{PRISM} approach,
in which the radial intracule densities and their derivatives are assembled
from fundamental integrals that involve modified spherical Bessel functions,
and they demonstrated that the Coulomb and exchange intracule densities $J(u)$
and $K(u)$ exhibit markedly different ranges: $J(u)$ is long-ranged and
comparable to the molecular extent, whereas $K(u)$ is short-ranged, with
delocalization effects introducing additional structure. The same algorithmic
philosophy was later extended to the Wigner intracule by Besley, O'Neill, and
Gill \cite{Besley2003}, who computed position, momentum, and phase-space
intracules from Hartree-Fock wave functions using Gaussian basis sets, and to
excited states by Besley and Gill \cite{Besley2004}, who used the CIS wave
function to generate intracules for low-lying excited states of atoms and
molecules.

A complementary line of work has focused on the connection between the
intracule and the energy. Since the intracule is the expectation value of a
two-electron operator, it is amenable to the energy-derivative formalism.
Cioslowski and Liu \cite{Cioslowski1998} combined this formalism with their
fast algorithm to obtain approximate intracule densities and Coulomb holes from
energy-derivative two-electron reduced density matrices, and they demonstrated
that routine calculations of correlated $I(\bm{R})$ are feasible at any level
of electronic structure theory. Their work also highlighted the importance of
the normalization sum rules and of the decomposition of the electron-pair
density into Coulomb, Dirac-Fock, and correlation contributions, which are
precisely the quantities that we use in the present work to validate our
formulation.

From a computational standpoint, the quantity of interest in most applications
is not the full vectorial density $I(\bm{R})$ but its spherical average, the
radial intracule density $I(R)$. Being a function of a single variable, $I(R)$
is far more amenable to visualization and interpretation, and its moments are
directly connected to physically meaningful quantities: the zeroth moment
reproduces the electron-pair count $N(N-1)$, while one half of the $(-1)$-moment yields the
electron-electron repulsion energy $V_{ee}$, and analogous relations hold for
the Coulomb and exchange-correlation components of the intracule. This close
connection between $I(R)$ and energetic quantities makes an accurate and
efficient evaluation of the radial intracule density highly desirable.

In practice, however, the evaluation of $I(R)$ is far from trivial. When the
wavefunction is expanded in a basis of Cartesian Gaussian functions (CGFs), the
almost universal choice in molecular electronic-structure calculations, the
vectorial intracule density can be written as a sum of four-index contributions
$I_{abcd}(\bm{R})$, each of which involves the product of two charge
distributions $\Omega_{ab}(\bm{r}_1)$ and $\Omega_{cd}(\bm{r}_2)$. The angular
average of these contributions has traditionally been performed numerically, for
instance on Lebedev-Laikov grids \cite{Lebedev1999}. This introduces a substantial
computational cost since, for each $\bm{R}$, $I(\bm{R})$ must be evaluated at
$N_{\theta\phi}$ points, and it also introduces a numerical error that is difficult to control,
particularly when high accuracy is required. Moreover, the numerical angular
integration must be repeated for every value of $\bm{R}$, which severely limits the
size of the systems that can be treated. Nevertheless, it should be acknowledged
that very clever numerical strategies have recently been proposed to carry out
the angular integration in an efficient and accurate form \cite{ylla-matito-ramoscordoba-2026}.
These authors developed multicenter integration schemes based on Becke's original proposal
\cite{Becke1988a} and on the topological fuzzy Voronoi cell
formalism \cite{Salvador2013}, opening the door to the systematic evaluation of
 intracule-based quantities in extended systems \cite{ylla-matito-ramoscordoba-2026}.
It is worth noting, in this respect, that Becke's monocentric integration scheme
is known to fail for DFT integrals involving high-order
derivatives \cite{zalesny:19jctc,sitkiewicz2022},
which is one of the motivations for the multicentric strategy adopted
in ref~\citenum{ylla-matito-ramoscordoba-2026}.
In spite of these advances, a fully analytical and exact method for performing the angular average remains
highly desirable, since it would eliminate the quadrature error altogether,
remove the need to repeat the angular integration for every value of $R$,
and provide closed-form expressions for the radial moments of the intracule density.

In this work we develop a general analytical framework for the radial intracule
density of molecular systems described by CGF basis sets. The starting point is
the well-known expansion of the products $\Omega_{ab}(\bm{r}_1)$ and
$\Omega_{cd}(\bm{r}_2)$ in Hermite Gaussian functions \cite{Helgaker2000}, which allows the
six-dimensional integral defining $I_{abcd}(\bm{R})$ to be factorized into three
one-dimensional integrals, which means that, as in the method of Cioslowski and Liu,
the computational cost associated with the evaluation of $I(\bm{R})$ on large three-dimensional
orthohedral grids grows with the cubic root of the total number of points.
We then show that the angular average over the
orientation of $\bm{R}$ can be carried out analytically, without resorting to
numerical quadrature. The central difficulty is the presence of the coupling
factor $e^{2\mu\bm{R}\cdot\bm{T}}$, where $\bm{T}=\bm{Q}_{cd}-\bm{P}_{ab}$ is the
separation between the Hermite centres associated with the two charge
distributions and $\mu$ is a reduced exponent. This factor couples the radial
and angular variables and is absent in the atomic case, where all Gaussian
centres coincide.

To handle this coupling we exploit an expansion of $e^{2\mu\bm{R}\cdot\bm{T}}$
in modified spherical Bessel functions and real spherical harmonics, which
decouples the angular variables at the price of introducing an infinite sum
over the angular momentum $l$; this sum is shown to truncate rigorously at
$l = p_x+p_y+p_z$, where the $p$'s are the Cartesian powers arising from the
Hermite expansion. In addition to this analytical route, a purely numerical
angular quadrature is retained as an independent reference against which the
analytical results can be validated. Particular emphasis is placed on the
numerical stability of the analytical approach: the products of large powers
and exponentials that arise in the molecular case are handled through iterative
rescaling and logarithmic representations, so that no intermediate quantity
overflows or underflows in double precision.

For atomic systems the general formalism simplifies dramatically. Since
$\bm{T}=\bm{0}$, only the $l=0$ term of the Bessel expansion survives, and the
radial intracule density reduces to a finite sum of terms of the form
$C_i(\mu,\zeta)\,R^{n_i}e^{-\mu R^2}$, with only even integer exponents $n_i$
ranging from 2 to $2+4l_{\max}$. This compact representation permits the
analytical evaluation of radial integrals, including the normalization integral
$\int_0^\infty I(R)\,dR$ and the electron-electron repulsion integral
$\int_0^\infty R^{-1}I(R)\,dR$, in terms of the Gamma functions. The same
simplification applies to any quartet of CGFs centred on a single atom of a
polyatomic molecule, so that the molecular and atomic formulations are connected
smoothly. It is worth emphasizing that the exact integration of $I(R)$, or of
$R^{-1}I(R)$, between zero and infinity is also possible in the general
molecular case. In this situation, however, the resulting integrals are no
longer expressible in terms of elementary functions and Gamma functions alone,
but are given in closed form in terms of the confluent hypergeometric function
${}_{1}F_{1}$.

A natural strategy to validate the formulation is to compare the values of
these radial integrals, obtained with the analytical approach described above,
and the associated energetic components, namely the Coulomb, exchange-correlation,
and Dirac-Fock contributions, against reference values obtained from independent
sources. This comparison is informative because the reference values provide an
independent route to the same quantities. It should be kept in mind, however,
that in certain regimes the analytical evaluation of these integrals is itself
affected by numerical errors. This occurs, for instance, for basis sets of CGFs
containing exponents of very large magnitude, or for very
large values of the radial variable $R$, where the analytical formulas involve
products of large powers and exponentials whose numerical evaluation in finite
precision may become unreliable. In such cases the comparison between the two
routes must be interpreted as a consistency check rather than as an absolute
error measure, and the observed discrepancies may reflect the limitations of
either method.

The remainder of this paper is organized as follows. In Sec.~\ref{sec:method} we recall the
definition of the intracule density and its expression in terms of the
second-order density matrix, and we derive the factorized form of
$I_{abcd}(\bm{R})$ in the Hermite Gaussian representation. Section~\ref{sec:methodradial} is devoted
to the radial intracule density: after deriving the general expression for
$I(R)$, we present the Bessel-function expansion (Sec.~\ref{sec:methodradial}) and the
numerical quadrature used for verification (Sec.~\ref{lebedev-laikov}). The atomic limit is
discussed in Sec.~\ref{atomic_intradule}, where the closed-form expression for $I_{abcd}(R)$ is
obtained. Section~\ref{sec:integrationintracule} describes the analytical integration over the radial
coordinate, expressing the result in closed form in terms of the confluent
hypergeometric function. Illustrative results for selected atomic and molecular
systems are presented in Sec.~\ref{sec:results}, after a discussion of the
numerical precision of the implementation in Sec.~\ref{sec:practical}, and the
main conclusions are summarized in Sec.~\ref{sec:conclusions}.

\section{The intracule density}
\label{sec:method}
Let $\Psi(1,N)$ be an arbitrary $N-$electron wavefunction and $\rho^2(\bm{r}_1,\bm{r}_2)$
its diagonal second-order density matrix (2-RDM),
\begin{equation}
\label{eq:2rdmdef}
\rho^2(\bm{r}_1,\bm{r}_2)=N(N-1)\int \vert \Psi(1,N) \vert^2 d\sigma_1 d\sigma_2 d\bm{x}_{i>2},\nonumber
\end{equation}
where $\bm{x}_i=\bm{r}_i\sigma_i$ represents the spatial-spin coordinates of electron $i$.
The intracule probability density (or simply intracule density) $I(\bm{R})$ is defined as
\begin{equation}
\label{eq:intra}
I(\bm{R}) = \int \rho^2(\bm{r}_1,\bm{r}_2)\, \delta(\bm{R}-\bm{s})\,d\bm{r}_1 d\bm{r}_2,
\end{equation}
where $\bm{s}=\bm{r}_2-\bm{r}_1$. $\rho^2(\bm{r}_1,\bm{r}_2)$ can be expanded
in a basis of CGFs,
\begin{equation}
\label{eq:defrho2}
\rho^2(\bm{r}_1,\bm{r}_2) =
\sum_{abcd} \gamma_{abcd} \phi_a^\star(\bm{r}_1) \phi_b(\bm{r}_1) \phi_c^\star(\bm{r}_2) \phi_d(\bm{r}_2).
\end{equation}
Then, $I(\bm{R})\equiv I(X,Y,Z)$ becomes
\begin{equation}
\label{eq:rho2}
	I(\bm{R}) =\sum_{abcd} \gamma_{abcd} I_{abcd}(\bm{R}),\hspace{0.4cm}\textrm{where}
\end{equation}
\begin{equation}
\label{eq:Iabcd}
I_{abcd}(\bm{R}) = \iint \Omega_{ab}(\bm{r}_1)\, \Omega_{cd}(\bm{r}_2)
\,\delta[\bm{R}-\bm{s}]\, d\bm{r}_1d\bm{r}_2,
\end{equation}
where $\Omega_{ab}(\bm{r}_1)$ and
$\Omega_{cd}(\bm{r}_2)$ are given by $\phi_a^\star(\bm{r}_1) \phi_b(\bm{r}_1)$
and $\phi_c^\star(\bm{r}_2) \phi_d(\bm{r}_2)$, respectively.

Before going further, it is worth noting that not all the quartets $(a,b,c,d)$
contribute significantly to the intracule density. Since the charge
distributions $\Omega^2_{ab}(\bm{r})$ and $\Omega^2_{cd}(\bm{r})$ are, by
construction, non-negative, the integrals
\begin{equation}
\int \Omega_{ab}^{\,2}(\bm{r})\,d\bm{r} > 0,
\qquad
\int \Omega_{cd}^{\,2}(\bm{r})\,d\bm{r} > 0
\end{equation}
provide a natural measure of the spatial extent and overlap of the two charge
distributions associated with a given quartet. Whenever one or both of these
quantities fall below a suitable threshold, the corresponding $I_{abcd}(\bm{R})$
contribution is negligible over the whole range of the interelectronic distance
$R$, and the quartet can therefore be omitted from the calculation altogether.
This simple screening criterion leads to substantial savings in computational
effort, particularly for large basis sets, without introducing any appreciable
loss of accuracy in the resulting radial intracule density.

Once all the $I_{abcd}(\bm{R})$ elements over CGFs have been obtained,
the usual four-index transformation can be carried out and $I(\bm{R})$ be expressed as
\begin{equation}
\label{eq:rho2mo}
I(\bm{R}) =\sum_{ijkl} \Gamma_{ijkl} I_{ijkl}(\bm{R}),
\end{equation}
where $\Gamma_{ijkl}$ and $I_{ijkl}(\bm{R})$ are the 2-RDM coefficients
and intracule densities in the MO basis ($\chi_i$), respectively.
The CGFs are given by
\begin{align}
\label{eq:cgtfa}
	\phi_a (\bm{r}_1,a,\bm{A}) & = (x_1^A)^{i}  (y_1^A)^{k}  (z_1^A)^{m}\;  e^{-a r_A^2}, \\
\label{eq:cgtfb}
	\phi_b (\bm{r}_1,b,\bm{B}) & = (x_1^B)^{j}  (y_1^B)^{l}  (z_1^B)^{n}\;  e^{-b r_B^2}, \\
\label{eq:cgtfc}
	\phi_c (\bm{r}_2,c,\bm{C}) & = (x_2^C)^{i'}  (y_2^C)^{k'}  (z_2^C)^{m'}  e^{-c r_C^2}, \\
\label{eq:cgtfd}
	\phi_d (\bm{r}_2,d,\bm{D}) & = (x_2^D)^{j'}  (y_2^D)^{l'} (z_2^D)^{n'}  e^{-d r_D^2},
\end{align}
where $\bm{A}\equiv(A_x,A_y,A_z)$, $\bm{B}\equiv(B_x,B_y,B_z)$, $\bm{C}\equiv(C_x,C_y,C_z)$,
and $\bm{D}\equiv(D_x,D_y,D_z)$ are the vector position of the
expansion centers, $x_1^A=x_1-A_x$, $\dots$,
$r_A^2=(x_1^A)^2+(y_1^A)^2+(z_1^A)^2$, $\dots$, and
$a$, $b$, $c$, and $d$ are CGF exponents.
The sums $i+k+m$, $j+l+n$, $i'+k'+m'$, and $j'+l'+n'$
define the angular momenta $l_a$, $l_b$, $l_c$, and $l_d$ of
$\phi_a$, $\phi_b$, $\phi_c$,  and $\phi_d$, respectively.
Expanding $\Omega_{ab}(\bm{r}_1)$ in terms of Hermite
Gaussian Functions,$\Lambda_{tuv}$ \cite{Helgaker2000} one has
\begin{equation}
\label{eq:omegaa}
\Omega_{ab}(\bm{r}_1) =
\sum_{t=0}^{i+j} \sum_{u=0}^{k+l} \sum_{v=0}^{m+n} E_t^{ij} E_u^{kl} E_v^{mn}
\Lambda_{tuv} (\bm{r}_1,p,\bm{P}_{ab}),
\end{equation}
\begin{equation}
\label{eq:defhgto}
\Lambda_{tuv} (\bm{r}_1,p,\bm{P}_{ab}) =
\frac{\partial^t}{\partial P_x}
\frac{\partial^u}{\partial P_y}
\frac{\partial^v}{\partial P_z}
\exp\left(-p r_P^2\right)
\end{equation}
where $p=a+b$, $\bm{P}_{ab}=(P_x,P_y,P_z)=(a\bm{A}+b\bm{B})/p$, and
$\bm{r}_P=\bm{r}_1-\bm{P}_{ab}$.
The expansion coefficients
$E_t^{ij}$, $E_u^{kl}$, and $E_v^{mn}$
can be obtained by recurrence expresions \cite{Helgaker2000} from
their $E_0^{00}$ values,
$\exp(-\mu_{ab} X_{AB}^2)$,
$\exp(-\mu_{ab} Y_{AB}^2)$, and
$\exp(-\mu_{ab} Z_{AB}^2)$, respectively,
with $\mu_{ab}=ab/p$. The product of these
three factors is $\exp\left(-\mu_{ab} R_{AB}^2\right)$, where $\bm{R}_{AB}=\bm{A}-\bm{B}$
and $R_{AB}=|\bm{R}_{AB }|$ is the distance between
centers $A$ and $B$. In a similar form,
\begin{equation}
\label{eq:omegab}
\Omega_{cd}(\bm{r}_2) =  \sum_{t'=0}^{i'+j'} \sum_{u'=0}^{k'+l'}
\sum_{v'=0}^{m'+n'} E_{t'}^{i'j'} E_{u'}^{k'l'} E_{v'}^{m'n'} \times
\Lambda_{t'u'v'} (\bm{r}_2,q,\bm{Q}_{cd}),
\end{equation}
where
\begin{equation}
\label{eq:defhgtobis}
\Lambda_{t'u'v'} (\bm{r}_2,q,\bm{Q}_{cd}) =
\frac{\partial^{t'}}{\partial Q_x}
\frac{\partial^{u'}}{\partial Q_y}
\frac{\partial^{v'}}{\partial Q_z}
\exp\left(-q r_Q^2\right)
\end{equation}
$q=c+d$, $\bm{Q}_{cd}=(Q_x,Q_y,Q_z)=(c\bm{C}+d\bm{D})/q$, and $\bm{r}_Q=\bm{r}_2-\bm{Q}_{cd}$.
The product of the exponential factors
$\exp\left(-p r_P^2\right)$ and
$\exp\left(-q r_Q^2\right)$ in Eqs.~\ref{eq:defhgto} and \ref{eq:defhgtobis},
respectively, can be written in the form
\begin{eqnarray}
\label{otra0}
F_{abcd} & = & \exp\left(-p r_P^2\right) \times \exp\left(-q r_Q^2\right) \\
\label{eq:otra1}
& = & \exp\left[ -\zeta (\bm{r}-\bm{S})^2 \right] \times \exp\left[-\mu \bm{F}^2\right],
\end{eqnarray}
where we have defined $\bm{r}=(1/2)(\bm{r}_1+\bm{r}_2)$,
$\zeta=p+q$, $\mu=pq/(p+q)$, $\eta = (p-q)/[2(p+q)]$,
$\bm{F}=(F_x,F_y,F_z)=\bm{s}-(\bm{Q}_{cd}-\bm{P}_{ab})=\bm{s}-\bm{T}$,
and
\begin{equation}
\label{eq:sotra}
\bm{S} =\frac{p\bm{P}_{ab}+q\bm{Q}_{cd}}{p+q}+ \eta\bm{s}.
\end{equation}
Multiplying Eqs.~\ref{eq:omegaa} and \ref{eq:omegab},
using Eqs.~\ref{eq:otra1}-\ref{eq:sotra},
$d\bm{r}_1 d\bm{r}_2= d\bm{r} d\bm{s}$,
and integrating over $\bm{r}$ and $\bm{s}$,
\begin{align}
\label{eq:Iabcd_factorized}
I_{abcd}(\bm{R}) & = I^x_{abcd}(X)\,I^y_{abcd}(Y)\,I^z_{abcd}(Z), \\
\label{eq:Ix_definition}
I^x_{abcd}(X) & = \sum_{t=0}^{i+j} \sum_{t'=0}^{i'+j'} E_{t}^{ij} E_{t'}^{i'j'}
\frac{\partial^t}{\partial P_x}
\frac{\partial^{t'}}{\partial Q_x} J^x_{abcd}(X), \\
\label{eq:jxabcd}
J^x_{abcd}(X) & = \int_{-\infty}^{+\infty} \exp\left[-\zeta (x-S_x)^2\right] dx\,
\times \int_{-\infty}^{+\infty} \delta(X-s_x)\, \exp\left[-\mu F_x^2\right] ds_x.
\end{align}
The first and second integrals of Eq.~\ref{eq:jxabcd} are $\left(\pi/\zeta\right)^{1/2}$
and $\exp\left(-\mu (X-T_x)^2\right)$, so that
\begin{equation}
\label{eq:Jx_result}
J^x_{abcd}(X) = \left(\frac{\pi}{\zeta}\right)^{1/2} \exp\left(-\mu F_{x}^2\right).
\end{equation}
Calling $J_{abcd}(\bm{R})=J^x_{abcd}(X)\,J^y_{abcd}(Y)\,J^z_{abcd}(Z)$,
the following compact expression for
$I_{abcd}(\bm{R})$ is obtained
\begin{equation}
\label{eq:final_Iabcd}
I_{abcd}(\bm{R}) = \sum_{tuv,t'u'v'}
E_{t}^{ij} E_{u}^{kl} E_{v}^{mn}
E_{t'}^{i'j'} E_{u'}^{k'l'} E_{v'}^{m'n'}\,
\times \frac{\partial^{tuv}}{\partial\bm{P}_{ab}}
\frac{\partial^{t'u'v'}}{\partial\bm{Q}_{cd}}
J_{abcd}(\bm{R}).
\end{equation}
When $l_a=l_b=l_c=l_d=0$,
a single term ($t=u=v=t'=u'=v'=0$) occurs in Eq.~\ref{eq:final_Iabcd}, with
$E_{t}^{ij} E_{u}^{kl} E_{v}^{mn}=\exp(-\mu_{ab} R_{AB}^2)$ and
$E_{t'}^{i'j'} E_{u'}^{k'l'} E_{v'}^{m'n'}=\exp(-\mu_{cd} R_{CD}^2)$.
Moreover, no derivatives of $J_{abcd}(\bm{R})$ are involved, so that
$I_{abcd}(\bm{R}) = K \exp(-\mu F^2)$,
where $K=(\pi/\zeta)^{3/2}\,\exp(-\mu_{ab} R_{AB}^2-\mu_{cd} R_{CD}^2)$,
which coincides with that given by Thakkar \textit{et al.}\ \cite{Thakkar1984a,Thakkar1984b}.
The spherically averaged (or radial) intracule density, defined as
$I_{abcd}(R)= R^2\int I_{abcd}(\bm{R})\, d\Omega$,
where $d\Omega=\sin(\theta) d\theta d\phi$ and
$\theta$ and $\phi$ are the spherical polar angles of $\bm{R}$,
is then given by
\begin{equation}
I_{abcd}(R)=4\pi\,\alpha^{-1} KR^2 e^{-\mu(R^2+T^2)} \sinh(\alpha),
\end{equation}
where $\alpha=2\mu R\vert \bm{T}\vert = 2\mu RT$, that coincides
with the result derived by Boyd \textit{et al.}\ \cite{Boyd1988}.
The closed-form expression above may be used instead of the general procedure
described below, provided that the numerical precautions discussed in the
Supporting Information (SI) are taken into account.

We now consider the general case. Substituting Eq.~\eqref{eq:Jx_result} into
Eq.~\eqref{eq:Ix_definition} for $I^x_{abcd}(X)$, defining $\xi_x=\sqrt{\mu}F_x$
with $F_x = X - T_x$ (and analogously for $F_y$ and $F_z$), using the identity
\begin{equation}
\label{eq:derivp}
 \frac{\partial \exp\left(-\mu F_x^2\right) }{\partial P_x}  =
-\frac{\partial \exp\left(-\mu F_x^2\right) }{\partial Q_x} =
\frac{\partial \exp\left(-\mu F_x^2\right) }{\partial F_x},
\end{equation}
and the Rodrigues formula for the Hermite polynomials,
\begin{equation}
\label{eq:hermite}
H_n(\xi)=(-1)^n\exp\left(\xi^2\right) \frac{d^n\exp\left(-\xi^2\right)}{d\xi^n},
\end{equation}
into Eq.~\ref{eq:Ix_definition} yields
\begin{equation}
\label{eq:Ix_Hermite}
I^x_{abcd}(X) = \sum_{n_x=0}^{l_{abcd}} A_{n_x}^{abcd}(\mu,\zeta) \exp\left(-\xi_x^2\right) H_{n_x}(\xi_x),
\end{equation}
with
\begin{equation}
\label{eq:A_coefficient}
A_{n_x}^{abcd}(\mu,\zeta)=\sqrt{\frac{\pi}{\zeta}} \mu^{n_x/2}
        \sum_{t+t'=n_x} (-1)^t E_t^{ab} E_{t'}^{cd}.
\end{equation}
The upper limit $l_{abcd}=l_a+l_b+l_c+l_d$ in Eq.~\eqref{eq:Ix_Hermite} is reached
only when $i=l_a$, $j=l_b$, $i'=l_c$, and $j'=l_d$.

The function $I^x_{abcd}(X)$ depends on $X$ only through the term
$\exp(-\xi_x^2)\,H_{n_x}(\xi_x)$. For fixed $\phi_a$, $\phi_b$, $\phi_c$,
and $\phi_d$, only this term must be recomputed when $X$ changes.
Moreover, the Hermite polynomials $H_{n_x}$ with $n_x>1$ follow from
$H_0(\xi_x)=1$ and $H_1(\xi_x)=2\xi_x$ through the well-known recurrence
relation
\begin{equation}
H_{n_x+1}(\xi_x)=2\xi_x\,H_{n_x}(\xi_x)-2n_x\,H_{n_x-1}(\xi_x).
\end{equation}
Derivatives of $I_{abcd}(\bm{R})$ of arbitrary order can be obtained at
almost no additional cost, using further recurrence relations satisfied by
the Hermite polynomials. We use the identity
\begin{equation}
H'_{n_x}(x)=2x\,H_{n_x}(x)-H_{n_x+1}(x),
\end{equation}
together with $\xi_x=\sqrt{\mu}\,(X-Q_x+P_x)$, and define
$f_{n_x}(\xi_x)=\exp(-\xi_x^2)\,H_{n_x}(\xi_x)$. We thus obtain
\begin{equation}
\frac{d f_{n_x}(\xi_x)}{dX}
= \sqrt{\mu}\,\frac{d f_{n_x}(\xi_x)}{d\xi_x}
= -\sqrt{\mu}\,f_{n_x+1}(\xi_x),
\end{equation}
and
\begin{equation}
\label{eq:Ix_first_deriv}
I^x_{abcd}(X)^{(1)} =  \frac{d I^x_{abcd}(X)}{dX}
 = \sum_{n_x=0}^{l_{abcd}} B_{n_x}^{abcd}(\mu,\zeta)\,
\exp(-\xi_x^2)\,H_{n_x+1}(\xi_x),
\end{equation}
where $B_{n_x}^{abcd}(\mu,\zeta)$ has an expression analogous to that of
$A_{n_x}^{abcd}(\mu,\zeta)$ in Eq.~\eqref{eq:A_coefficient}, but with
$(-1)^{t+1}$ in place of $(-1)^t$ and $\mu^{(n_x+1)/2}$ in place of
$\mu^{n_x/2}$. The $n$-th derivative of $I^x_{abcd}(X)$ with respect to
$X$ is then
\begin{equation}
\label{eq:Ix_nth_deriv}
I^x_{abcd}(X)^{(n)}
= \sum_{n_x=0}^{l_{abcd}} C_{n_x}^{abcd}(\mu,\zeta)\,
\exp(-\xi_x^2)\,H_{n_x+n}(\xi_x),
\end{equation}
where $C_{n_x}^{abcd}(\mu,\zeta)$ is defined as $A_{n_x}^{abcd}(\mu,\zeta)$
but with $(-1)^{t+n}$ in place of $(-1)^t$ and $\mu^{(n_x+n)/2}$ in place
of $\mu^{n_x/2}$. The derivatives with respect to $Y$ and $Z$ are
completely analogous.

The evaluation of $I(\bm{R})$ is organized in four steps.

(1) The primitive CGFs of the system are grouped
into shells. Referring to $\phi_a$ in Eq.~\eqref{eq:cgtfa}, a shell is
defined as the set of $(l_a+1)(l_a+2)/2$ CGFs that share the same angular
momentum $l_a=i+k+m$ and the same exponent $a$. Formally, such a shell is
written as $\varphi_a=(x_A,y_A,z_A)^{l_a}\exp(-a r_A^2)$.

(2) For every pair of shells $(\varphi_a,\varphi_b)$, the coefficients
$E_{t}^{ij}$, $E_{u}^{kl}$, and $E_{v}^{mn}$ are computed by recurrence
from their corresponding $E_{0}^{00}$ values. The indices run over
$i,k,m=0,\dots,l_a$; $j,l,n=0,\dots,l_b$; $t=0,\dots,i+j$;
$u=0,\dots,k+l$; and $v=0,\dots,m+n$.

(3) Once these coefficients are available, the $A_{n_x}^{abcd}(\mu,\zeta)$
coefficients are computed by means of Eq.~\eqref{eq:A_coefficient}. Then,
for each grid point $(X,Y,Z)$, the exponential factors $\exp(-\xi_x^2)$ and
the Hermite polynomials $H_{n_x}(\xi_x)$, $H_{n_y}(\xi_y)$, and
$H_{n_z}(\xi_z)$ are evaluated for all orders
$0\le n_x,n_y,n_z\le l_{abcd}$.

(4) Finally, Eq.~\eqref{eq:Ix_Hermite} and its counterparts along $Y$ and
$Z$ are combined to obtain the full integral.

The computational cost of evaluating $I(\bm{R})$ for an arbitrary set of
$N_{\bm{R}}$ points scales linearly with $N_{\bm{R}}$. Nevertheless, this
cost can be substantially reduced when the points form a Cartesian
grid. In that case, if the grid contains $N_{Rx}$ distinct $X$ coordinates,
$N_{Ry}$ distinct $Y$ coordinates, and $N_{Rz}$ distinct $Z$ coordinates,
one may precompute and store, either in memory or on disk, all the
one-dimensional integrals $I_{abcd}^x(X)$, $I_{abcd}^y(Y)$, and
$I_{abcd}^z(Z)$. These precomputed values can then be reused for each
$\bm{R}=(X,Y,Z)$, reducing the overall scaling to approximately
$3N_{\bm{R}}^{1/3}$. This is precisely the scaling reported by Cioslowski
and Liu for the evaluation of intracule and extracule densities on large
grids of points \cite{Cioslowski1996}.

\section{The radial intracule density}
\label{sec:methodradial}
In the vast majority of applications of the intracule density, the quantity
of interest is not the full vectorial function $I(\bm{R})$ but its spherical
average $I(R)$, defined as
\begin{equation}
\label{eq:irdef}
I(R)=R^2\int I(\bm{R})\,d\Omega = \sum_{abcd} \gamma_{abcd}\,I_{abcd}(R),
\end{equation}
with
\begin{equation}
\label{eq:irdefabcd}
I_{abcd}(R)=R^2\int I_{abcd}(\bm{R})\,d\Omega.
\end{equation}
Being a function of a single variable, $I(R)$ is far more manageable than
$I(\bm{R})$, yet its detailed analysis provides a great deal of information
about the electronic correlation in a given molecular system.

Traditionally, the angular average has been evaluated numerically, for
instance on a single-center Lebedev--Laikov (SCLL) $(\theta,\phi)$ grid
\cite{Lebedev1999} placed at the origin ($\bm{R}=\bm{0}$). Only recently has
a more refined strategy been proposed \cite{ylla-matito-ramoscordoba-2026}:
instead of relying on a single SCLL grid, $M+1$ such grids are combined,
where $M$ is the number of atomic pairs in the molecule (hydrogen atoms
excluded). One grid is centered at the origin, and the remaining $M$ are
centered at the points $\bm{R}_{AB}$ corresponding to each pair of distinct
atoms $A$ and $B$. This arrangement was designed to mitigate inaccuracies in
the computation of the intracule moments $\int_0^\infty R^n I(R)\,dR$.

For a SCLL grid, if the angular grid consists of $N_{\theta\phi}$ points,
computing $I(R)$ for $N_R$ values of $R$ requires the evaluation of
$I(\bm{R})$ at $N_{R\theta\phi}=N_R\times N_{\theta\phi}$ points $\bm{R}$,
which entails a considerable computational effort. For this reason, and also
with the aim of minimizing the errors associated with the numerical
integration and avoiding any loss of precision, it is highly desirable to
carry out the angular average of $I(\bm{R})$ in as analytical a way as
possible. This is precisely the purpose of this section. We consider the
general case of the evaluation of $I(\bm{R})$ for an arbitrary molecule and
show that, in the particular case of an atomic system (where all CGFs are
centered on the single existing nucleus), the general expression simplifies
enormously and reduces to a relatively simple analytical expression.

Our starting point is Eqs.~\eqref{eq:Iabcd_factorized} and \eqref{eq:Ix_Hermite}. With the definitions $(F_x,F_y,F_z)=(X-T_x,Y-T_y,Z-T_z)$ and $(\xi_x,\xi_y,\xi_z)=\mu^{1/2}(X-T_x,Y-T_y,Z-T_z)$, which imply
\begin{equation}
\exp\!\bigl(-(\xi_x^2+\xi_y^2+\xi_z^2)\bigr)
=
\exp\!\bigl(-\mu(\bm{R}-\bm{T})^2\bigr),
\end{equation}
and after substituting Eq.~\eqref{eq:Ix_Hermite} into Eq.~\eqref{eq:Iabcd_factorized}, the latter can be recast as
\begin{equation}
\label{eq:rad1}
I_{abcd}(\bm{R})  =  \sum_{n_x,n_y,n_z=0}^{l_{abcd}} A_{n_x n_y n_z}^{abcd}(\mu,\zeta)\,
\times H_{n_x}(\xi_x) H_{n_y}(\xi_y) H_{n_z}(\xi_z) \times e^{-\mu(\bm{R}-\bm{T})^2},
\end{equation}
where $A_{n_x n_y n_z}^{abcd}(\mu,\zeta)=A_{n_x}^{abcd}(\mu,\zeta)\times A_{n_y}^{abcd}(\mu,\zeta) \times A_{n_z}^{abcd}(\mu,\zeta)$.
The Hermite polynomial $H_{n_x}(\xi_x)$ in powers of $\xi_x$ is given by $H_{n_x}(\xi_x) =
\sum_{i_x=0}^{n_x} C^{x}_{i_x} \xi_x^{i_x}$, where the $C^{x}_{i_x}$ coefficients are known.
However, for the present purpose it is more convenient to put in powers of $X$
\begin{equation}
\label{eq:thermi}
H_{n_x}(\xi_x) = \sum_{i_x=0}^{n_x} D^{x}_{i_x} X^{i_x},
\end{equation}
with $D^{x}_{i_x}=\sum_{k=i_x}^{n_x} C^{x}_{k} \times {\mu}^{k/2} \binom{k}{i_x} (-T_x)^{k-i_x}$.
Analogue expressions can be used for $H_{n_y}(\xi_y)$ and $H_{n_z}(\xi_z)$. Equation \ref{eq:rad1}
can then be rewritten as
\begin{equation}
\label{eq:rad2}
I_{abcd}(\bm{R}) = \sum_{p_x,p_y,p_z=0}^{l_{abcd}} D_{p_x p_y p_z}^{abcd}(\mu,\zeta)
X^{p_x} Y^{p_y} Z^{p_z}\, \times e^{-\mu(R^2+T^2)} \,e^{2\mu\bm{R}\cdot\bm{T}},
\end{equation}
where the factor $e^{-\mu (\bm{R} - \bm{T})^2}$ has been written as
\begin{equation}
\label{eq:exp_factor_molecular}
e^{-\mu (\bm{R} - \bm{T})^2} = e^{-\mu(R^2+T^2)} e^{2\mu \bm{R}\cdot\bm{T}},
\end{equation}
\begin{equation}
\label{2sumprim}
D_{p_x}^{abcd} =
\sum_{n_x=p_x}^{l_{abcd}} A_{n_x}^{abcd}(\mu,\zeta)\, D^{x}_{p_x},
\end{equation}
and
\begin{equation}
\label{Dpxpypz}
D_{p_x p_y p_z}^{abcd}(\mu,\zeta)=D_{p_x}^{abcd}\times D_{p_y}^{abcd} \times D_{p_z}^{abcd}.
\end{equation}
The exponential factor $e^{2\mu\bm{R}\cdot\bm{T}}$ in Eq.~\ref{eq:rad2}
is now expanded in the form
\begin{equation}
\label{expfactor}
e^{2\mu\bm{R}\cdot\bm{T}} =\sum_{l=0}^{\infty} (2l+1)\,
i_{l}(\alpha)\, P_{l}(\cos{\gamma})
\end{equation}
where $\alpha=2\mu\vert\bm{R}\vert \vert\bm{T}\vert = 2\mu R T$, $i_{l}(\alpha)$ is the modified
spherical Bessel function of the first kind of integer order $l$, $P_{l}(\cos\gamma)$ is the Legendre
polynomial of degree $l$, and $\gamma$ is the angle between vectors $\bm{R}$ and $\bm{T}$.
Since $\bm{T}\equiv(T\sin\theta_T\cos\phi_T,T\sin\theta_T\sin\phi_T,T\cos\theta_T)$,
$\gamma$ mixes the angular variables $\theta$ and $\phi$ (from which $\bm{R}$ depends)
with $\theta_T$ and $\phi_T$, which are fixed quantities for each $\bm{T}$. To split them,
we expand $P_{l}(\cos{\gamma})$ as
\begin{equation}
\label{addharmoreal}
P_{l}(\cos{\gamma})=\frac{4\pi}{2l+1}\sum_{m=-l}^{l} S_{lm}(\theta,\phi)\,
S_{lm}(\theta_T,\phi_T),
\end{equation}
where $S_{lm}$ are the real spherical harmonics (RSH),
\begin{equation}
\label{rsh}
S_{lm} (\theta,\phi) = \Theta_{l|m|} (\theta)\, \Phi_m(\phi),\hspace{0.4cm} \textrm{with}
\end{equation}
\begin{equation}
\Theta_{lm} (\theta) =
N_{lm} P_l^m (\cos\theta),
\end{equation}
where $N_{lm}$ is a normalization constant and $P_l^m (x)$ are the associated
Legendre polynomials,
and
\begin{equation}
\Phi_m (\phi) = \left\{
    \begin{array}{ll}
      \sqrt{2} \cos m\phi   & m>0, \\
      1                     & m=0, \\
      \sqrt{2} \sin |m|\phi & m<0.
    \end{array} \right.
\label{funcionazimutal}
\end{equation}
Substituting Eq.~\ref{addharmoreal} in Eq.~\ref{expfactor}, Eq.~\ref{expfactor}
in Eq.~\ref{eq:rad2},
and integrating the angles,
\begin{equation}
\label{otheradbis}
I_{abcd}(R) = \sum_{p_x,p_y,p_z=0}^{l_{abcd}} D_{p_x p_y p_z}^{abcd}(\mu,\zeta)
R^{\,p_x+p_y+p_z+2}\,\times J_{p_xp_yp_z}(\mu,R,T),
\end{equation}
with
\begin{equation}
\label{angintj3}
J_{p_xp_yp_z}(\mu,R,T) = 4\pi\,\sum_{l=0}^{\infty} \sum_{m=-l}^{l}
\left[B_l(\mu,R,T) \times S_{lm}(\theta_T,\phi_T)\times G_{lm}(p_x,p_y,p_z) \right],
\end{equation}
where the overall factor $4\pi$ originates from the addition theorem
(Eq.~\ref{addharmoreal}) applied to the expansion of Eq.~\ref{expfactor},
\begin{equation}
\label{besselscaled}
	B_l(\mu,R,T) = e^{-\mu\left(R^2+T^2\right)}\,i_{l}(2\mu RT),
\end{equation}
\begin{equation}
\label{Jtheta}
G_{lm}(p_x,p_y,p_z) = \int_{0}^{\pi}\int_{0}^{2\pi}
	\hspace{-0.3cm} S_{lm}(\theta,\phi)\, g_{p_xp_yp_z}
d\theta d\phi,
\end{equation}
and
\begin{equation}
g_{p_xp_yp_z}=\sin^{p_x+p_y+1}(\theta) \cos^{p_z}(\theta)
\sin^{p_y}(\phi) \cos^{p_x}(\phi).\nonumber
\end{equation}
Neither the angular integrals~\eqref{Jtheta} nor the real spherical harmonics (RSHs) $S_{lm}(\theta_T,\phi_T)$ appearing in Eq.~\eqref{angintj3} depend on $R$. Consequently, the full $R$ dependence of $I_{abcd}(R)$ is carried by $B_{l}(\mu,R,T)$ and by the factor $R^{p_x+p_y+p_z+2}$ of Eq.~\eqref{otheradbis}. In addition, it is shown in Section~3 of the SI that $G_{lm}(p_x,p_y,p_z)=0$ whenever $l>p_x+p_y+p_z$, so that the sum over $l$ in Eq.~\eqref{angintj3} can safely be truncated at $l=p_x+p_y+p_z$. Although $m$ in Eq.~\eqref{angintj3} takes negative values when $l\ne 0$, the definition of the RSHs in Eqs.~\eqref{rsh} and \eqref{funcionazimutal} implies that $G_{lm}(p_x,p_y,p_z)$ never needs to be evaluated for $m<0$. The detailed calculation of $G_{lm}(p_x,p_y,p_z)$ is given in Section~3 of the SI.
At this point, however, it is important to stress that they can be obtained from the
basic angular integrals
\begin{equation*}
\Theta(a,b) = \int_0^{\pi} \sin^{a+1}(\theta) \cos^{b}(\theta) \, d\theta,\qquad
\Phi(c,d) = \int_0^{2\pi} \sin^{c}(\phi) \cos^{d}(\phi) \, d\phi.
\end{equation*}
Both can be recursively obtained from $\Theta(0,0)=2$ and $\Phi(0,0)=2\pi$.
Moreover, since $\Phi(c,d)$ vanishes unless $c$ and $d$ are both even, and
$\Theta(a,b)$ vanishes for odd $b$, the $p_x$, $p_y$, $p_z$ terms in Eq.~\ref{otheradbis}
need to be computed only for even values of these three indices (see Section~3 of the SI).

Finally, it is necessary to discuss in some detail the computation of $B_l(\mu,R,T)$
in Eq.~\ref{besselscaled}. Since $i_{l}(\alpha)$ tends to $\infty$ while
$\exp{[-\mu(R^2+T^2)]}$
tends to 0 for large values of $\alpha=2\mu RT$,
some care is needed in the evaluation of its product.
If both factors are computed separately in double precision, overflow
(the argument of the exponential easily exceeds $\approx 709$, the limit of
binary64) or underflow can occur, even when the final product is perfectly
representable. To avoid this, $B_l(\mu,R,T)$
is rewritten in the form
\begin{equation}
\label{besselscaledbis}
B_l(\mu,R,T) = e^{-\mu(R-T)^2} \, i_l^{\mathrm{scaled}}(\alpha)
\end{equation}
where $i_l^{\mathrm{scaled}}(\alpha) = e^{-\alpha}\,i_l(\alpha)$
is the scaled version of the modified spherical Bessel function,
which is provided, for instance, by
the GNU Scientific Library (GSL) \cite{gslteam2019gnuscientificlibrary}.
Since $\mu \ge 0$, the exponent $-\mu (R-T)^2$ is always $\le 0$, so
$e^{-\mu (R-T)^2}$ is bounded between 0 and 1 (never overflows), and
$i_l^{\mathrm{scaled}}(\alpha)$ is bounded (decaying as $1/(2\alpha)$
for large $\alpha$), so the product never overflows, regardless of the values of
$\mu$, $R$, and $T$.
Furthermore, if $\mu (R-T)^2 > 700$,
$B_l(\mu,R,T)$ is considered numerically negligible (its prefactor
$e^{-\mu (R-T)^2}$ is then bounded by $e^{-700}$) and is directly forced to 0.0
(avoiding unnecessary computations).
Finally, the result is also provided in logarithmic scale,
$\log[B_l(\mu,R,T)] = -\mu (R-T)^2 + \log[i_l^{\mathrm{scaled}}(\alpha)]$,
which is the recommended representation when $B_l$ is so small that it
underflows to 0.0 in double precision but its order of magnitude is
still needed.

\subsection{Numerical radial intracule density}
\label{lebedev-laikov}
A numerical angular integration to compute the radial density that bypasses the analytical
expansion of the coupling exponential and the use of modified spherical Bessel functions
has also been implemented.
Using $(X,Y,Z) = (R\sin\theta\cos\phi,R\sin\theta\sin\phi,R\cos\theta)$ in Eq.~\ref{eq:rad2}
and integrating over $(\theta,\phi)$, one obtains Eq.~\ref{otheradbis}, this time with
$J_{p_xp_yp_z}(R)$ given by
\begin{eqnarray}
\label{otherad5}
J_{p_xp_yp_z}(R) \!\!\!\!& = &
	\!\!\!\!\!\!\iint g_{p_xp_yp_z}\,
e^{-\mu(\bm{R}-\bm{T})^2}
d\theta d\phi.
\end{eqnarray}
The above angular integration is performed by a standard Lebedev-Laikov quadrature \cite{Lebedev1999}:
\begin{eqnarray}
\label{otherad6}
	\hspace{-0.5cm}J_{p_xp_yp_z}(R) \!\!\!\!& \simeq & \!\!\!\!\sum_i^{N_{\textrm{ang}}}
\omega_i(\theta_i,\phi_i)\, g(\theta_i,\phi_i)\, F(\theta_i,\phi_i),
\end{eqnarray}
where
$F(\theta_i,\phi_i)=e^{-\mu(R^2+T^2)}\,e^{2\mu \bm{R} \cdot \bm{T}}$,
with $\bm{R} \cdot \bm{T}= R [T_x \sin(\theta_i)\cos(\phi_i)+
T_y \sin(\theta_i)\sin(\phi_i) + T_z\cos(\theta_i)]$.

A useful test of the validity of the expressions derived in this Section is
to compare the integrals $\int_0^{\infty} R^n I(R) dR$ ($n=-1,0,1,\dots$)
against the value
$N(N-1)$ for $n=0$ or $V_{ee}$ for $n=-1$. Exact analytical expressions for these integrals
in terms of the confluent hypergeometric function ${}_{1}F_{1}$, are derived in
the Section \ref{sec:exact_radial}.

\subsection{Atomic radial intracule density}
\label{atomic_intradule}
The atomic case leads to drastic simplifications in the expression for
$I(R)$. For an atom, all CGFs are centered on the same nucleus, and
therefore the vector $\bm{T}=\bm{Q}_{cd}-\bm{P}_{ab}$ vanishes identically
for every quartet $abcd$. Moreover, it can be verified explicitly that the
coefficients $E_{t}^{ij}$ appearing in the Hermite Gaussian expansion of
$\Omega_{ab}(\bm{r})$, Eq.~\eqref{eq:omegaa}, are zero unless $t+i+j$ is
even, with $E_{0}^{00}=1$. The final expression for $I_{abcd}(\bm{R})$ then
follows simply by setting $\bm{T}=0$ in Eq.~\eqref{eq:rad2}.
\begin{equation}
\label{eq:rad2atom}
I_{abcd}(\bm{R}) =
\!\!\!\!\!\!\!\!\!\sum_{p_x,p_y,p_z=0}^{l_{abcd}} \!\!\!\!\!\!\!\!
D_{p_x p_y p_z}^{abcd}(\mu,\zeta) e^{-\mu R^2} \,X^{p_x} Y^{p_y} Z^{p_z}\nonumber.
\end{equation}
Integrating the angles,
\begin{equation}
\label{eq:Iabcd_radial}
I_{abcd}(R)  = \sum_{n_x,n_y,n_z=0}^{l_{abcd}} D^{abcd}_{n_xn_yn_z}(\mu,\zeta)
e^{-\mu R^2} \,\times R^{n_x+n_y+n_z+2} \Theta(n_x+n_y,n_z)\Phi(n_x,n_y).
\end{equation}
Combining the previous results, we obtain the following closed-form
expression for $I_{abcd}(R)$:
\begin{equation}
\label{eq:Iabcd_final_atom}
I_{abcd}(R) = \sum_i C_i(\mu,\zeta)\, R^{\,n_i}\, e^{-\mu R^2},
\end{equation}
where the index $i$ runs over the surviving triples $(n_x,n_y,n_z)$, the
exponents $n_i$ are positive even integers, and the coefficients
$C_i(\mu,\zeta)$ collect the products
$D^{\,abcd}_{n_xn_yn_z}(\mu,\zeta)\,\Theta(n_x+n_y,n_z)\,\Phi(n_x,n_y)$
introduced above.
Recalling Eqs.~\eqref{eq:rho2} and~\eqref{eq:irdef}, it follows that $I(R)$ can be cast in
a form analogous to Eq.~\eqref{eq:Iabcd_final_atom}, with the $C_i$ coefficients now depending on
all the exponents present in the basis set of CGFs. Moreover, the exponents $n_i$ assume
only even values, running from $2$ to $2+4l_{\mathrm{max}}$ in steps of 2,
where $l_{\mathrm{max}}$ denotes the maximum angular momentum number contained in the basis.
Of course, Eq.~\eqref{eq:Iabcd_radial} remains valid in the molecular case, as long as the quartets
$abcd$ under consideration have all CGFs centered on a single nucleus of the molecule.

The integration of $I_{abcd}(R)$ from 0 to $\infty$, that is considered in the following section for the general case, is particularly simple in atoms
\begin{equation}
\label{good}
\int_0^\infty I_{abcd}(R)\, dR =\sum_i C_i(\mu,\zeta)
\frac{\Gamma\left[(n_i+1)/2\right]}{2\mu^{(n_i+1)/2}}.
\end{equation}
The contribution of the quartet $I_{abcd}$ to the total
electron-electron repulsion energy is then given
by
\begin{equation}
\label{veeabcd}
V_{ee}^{abcd}=\frac{1}{2} \int_0^{\infty} R^{-1} I_{abcd}(R)\,dR =\sum_i C_i(\mu,\zeta)
\frac{\Gamma\left[n_i/2\right]}{4\mu^{n_i/2}}
\end{equation}

\section{Integration of the radial intracule density}
\label{sec:integrationintracule}
In this section we describe the three methods employed in this work to
evaluate the moments of the radial intracule density,
$\int_0^{\infty} R^n I(R)\,dR$, namely: (i) a numerical integration with
standard quadratures, combined with a mapping $R = f(u)$ that transforms
the semi-infinite interval $R\in[0,\infty)$ into an interval $u\in[a,b]$ with
finite lower and upper limits; (ii) an adaptive integration scheme using
a non-fixed number of points $R_i$, which concentrates the evaluations of
$I(R_i)$ where they are most needed in order to improve the accuracy of
the resulting moments; and (iii) an exact integration based on the
evaluation of the confluent hypergeometric function with different
arguments.

\subsection{Standard numerical quadrature}
The procedure is modelled on the strategy customarily employed in Density
Functional Theory (DFT) to evaluate $\int_{\mathbb{R}^3} F[\rho(\bm{r})]\,d\bm{r}$,
where $F[\rho(\bm{r})]\equiv F(\bm{r})$ is an arbitrary functional of the
electron density $\rho(\bm{r})$. Following Becke's proposal, $F(\bm{r})$ is
decomposed as a sum over the atoms of the molecule,
$F(\bm{r})=\sum_A \omega_A(\bm{r})\,F(\bm{r})=\sum_A F_A(\bm{r})$, where the
weight functions $\omega_A(\bm{r})$ satisfy $\sum_A \omega_A(\bm{r})=1$ at
every $\bm{r}$. The angular integration is then carried out (for instance on
the Lebedev--Laikov angular points $\theta_i,\phi_i$ discussed in
Sec.~\ref{lebedev-laikov}), giving
$F_A(r)=\int F_A(\bm{r})\sin\theta\,d\theta\,d\phi$. Finally, the radial
integration $\int_0^\infty F_A(r)\,r^2\,dr$, which is analogous to the one
required in this section, is performed after a variable transformation
$R=R(u)$ that maps the semi-infinite interval $[0,\infty)$ onto a finite
computational range $u\in[a,b]$. Various mappings $R(u)$ have been used over
the years \cite{Becke1988a,MuraKnowles1996,TreutlerAhlrichs1995,MurrayHandy1993,Lindh2001,Tehrani2024,KrackKoster1998},
differing in their numerical performance and in the way they
accumulate points in $[0,\infty)$. Our code for integrating $R^n I(R)$
implements two of them: the Euler--Maclaurin mapping \cite{MurrayHandy1993}
\[
R(u) = \alpha_A \frac{u^m}{(1-u)^m}, \qquad m=1,2,
\]
and the radial mapping originally proposed by Becke \cite{Becke1988a},
\[
R(u) = \alpha_A \frac{1+u}{1-u},
\]
with $u\in[0,1]$ and $u\in[-1,1]$, respectively. In both cases, $\alpha_A$ is
chosen so that roughly half of the original $R$ grid points lie below and
above the midpoint of the $u$ interval. In the DFT context, $\alpha_A$ is
typically taken as the Slater--Bragg radius of atom $A$.
Upon application of the mapping, the integral of $R(u)^n I(R(u))$
is evaluated by means of a simple trapezoidal (Euler--Maclaurin) rule and a Gauss--Legendre
quadrature \cite{Press_1992_NumericalRecipes}, respectively.

When the three-step strategy described above is used in DFT integrations, the
resulting integrals are usually fairly accurate, except
in integrals involving high-order derivatives \cite{zalesny:19jctc}.
One might expect the same to hold
for the radial integration of $R^n I(R)$. Unfortunately, this is not the case,
as recently shown by Ylla, Ugalde, Matito, and Ramos-Cordoba, who gave a
convincing explanation for this poor behaviour \cite{ylla-matito-ramoscordoba-2026}.
The function $I(R)$ has a first
maximum at some distance from the origin $R=0$ that depends on the molecule. In
a DFT context, for values of $R$ beyond this maximum, the integrand
$F[\rho(\bm{r})]$ usually decays smoothly as $R$ increases. This does not happen
here, since $I(R)$ generally exhibits maxima at values of $R$ corresponding to
many of the interatomic distances of the molecule under study. Ylla, Ugalde,
Matito, and Ramos-Cordoba proposed a solution that works well for improving the
accuracy of the moments of $I(R)$. In the following subsection we present an
alternative approach to the same end.

\subsection{Adaptive radial integration}
To avoid the loss of accuracy in the computation of the moments of $I(R)$, an
alternative to the proposal of Ylla, Ugalde, Matito, and Ramos-Cordoba consists
in abandoning the construction of a grid with a fixed number of values of $R$,
and instead increasing this number as much as necessary, reducing the
integration error below a threshold fixed in advance. At the same time, the
values of $R$ must accumulate in the regions where this error decreases most
effectively, which are usually those with the largest values of $R$ or those
where the curvature of $I(R)$ changes. In our work we have exploited this idea
through the one-dimensional adaptive integration implemented in the
\texttt{QUADPACK} package \cite{piessens1983quadpack}. More specifically, we have used the routine
\texttt{dqagi}, a general-purpose globally adaptive integrator with integration
limits $[a,\infty]$, $[-\infty,b]$, or $[-\infty,+\infty]$, designed to satisfy
the accuracy condition $|I_{\textrm{exact}} - I_{\textrm{approx}}| \le
\textrm{max}(\textrm{epsabs},\,\textrm{epsrel}\times |I_{\textrm{approx}}|)$,
where the absolute error and relative error tolerances ($\textrm{epsabs},\textrm{epsrel}$)
are externally given.

\subsection{Exact integration}
\label{sec:exact_radial}
In this subsection, we sketch the derivation of an analytical expression for
$\int_0^{\infty} R^n I(R) dR$. Full details are given in Section~4 of the SI.
Taking into account Eq.~\ref{eq:irdef} and the expressions
\ref{otheradbis}--\ref{Jtheta}, it is clear that the above integral can
be recast in terms of the simpler radial integrals
\begin{equation}
I(n,l,\mu,T) = \int_0^\infty R^{\,n}\,B_l(\mu,R,T) \, dR.
\label{inlmut}
\end{equation}
where $B_l(\mu,R,T)$ is given by Eq.~\ref{besselscaled} or Eq.~\ref{besselscaledbis}.
Using
\begin{equation}
i_l(z) = \sqrt{\frac{\pi}{2z}}\, I_{l+1/2}(z),
\label{eq:spherical_bessel}
\end{equation}
where $I_{\nu}(z)$ is the standard modified Bessel function, $B_l(\mu,R,T)$ becomes
\begin{equation}
\label{eq:Bl_final}
B_l(\mu,R,T) = \sqrt{\frac{\pi}{4\mu T}}\,
	e^{-\mu(R^2+T^2)} \frac{I_{l+\nicefrac{1}{2}}(2\mu RT)}{R^{\nicefrac{1}{2}}}.
\end{equation}
Substituting Eq.~\ref{eq:Bl_final} into Eq.~\ref{inlmut}:
\begin{equation}
\label{eq:Inl_integral}
I(n,l,\mu,T)  = \sqrt{\frac{\pi}{4\mu T}}\, e^{-\mu T^2}\,
\times \int_{0}^{\infty} R^{n-\nicefrac{1}{2}} e^{-\mu R^2}
I_{l+1/2}(2\mu RT)\,dR.
\end{equation}
A closed form for the above integral is given by
Eq.~6.631.1 of Gradshteyn \& Ryzhik book \cite{gradshteyn2014table}.
The final expression for $I(n,l,\mu,T)$ is
\begin{eqnarray}
\label{maineq:hypergeom}
I(n,l,\mu,T) & = &
\frac{\sqrt{\pi}}{4}\; e^{-\mu T^2}\,
	\mu^{(l-n-1)/2}\, T^{\,l}\,
\frac{\Gamma\!\left(\tfrac{n+l+1}{2}\right)}
     {\Gamma\!\left(l+\tfrac{3}{2}\right)} \nonumber\\
	&\times& {}_1F_1\!\left(\frac{n+l+1}{2};\, l+\tfrac{3}{2};\, \mu T^2\right),
\end{eqnarray}
with the $T=0$ special case
\begin{equation}
I(n,0,\mu,0) \;=\; \tfrac{1}{2}\,\mu^{-(n+1)/2}\,
\Gamma\!\left(\tfrac{n+1}{2}\right),
\end{equation}
which vanishes for $l>0$ since $i_{l\ne 0}(0)=0$. A full derivation
of Eq.~\ref{maineq:hypergeom} is given in Section~4 of the SI.
For the norm, $n-l$ is even and ${}_1F_1$ reduces to a terminating
polynomial (Sec.~\ref{sec:practical}). For the $R^{-1}$ moment that gives
$V_{ee}$, $n-l$ is odd, $a$ is a positive integer and $b$ a
half-integer, and ${}_1F_1$ becomes an error-function (Boys-type)
expression that no longer terminates; its evaluation is described in
Section~S4.1 of the SI.
Some care is needed to evaluate $_1F_1$ depending on the values
of its arguments. Section 4 of the SI discusses
numerically stable methods for this purpose.
Just as with $i_l(z)$, it is sometimes more convenient to
evaluate the exponentially damped product $e^{-z}\,{}_1F_1(a,b,z)$,
which is the quantity that actually appears in Eq.~\ref{maineq:hypergeom} if $z$ is identified
with $\mu T^2$.

\section{Practical aspects of the numerical implementation}
\label{sec:practical}
The analytical formulation of the preceding sections is exact in
principle: the coupling table $J_{p_xp_yp_z}$ of Eq.~\eqref{angintj3}
and its contraction with the $A$-coefficients of
Eq.~\eqref{eq:A_coefficient} are finite, exact polynomial identities,
and no truncation error is introduced anywhere in the method. The only
failure mode is therefore arithmetic, and it manifests itself in two
ways: catastrophic cancellation in the angular-momentum coupling, and
overflow or underflow in the evaluation of the coupling functions.
Both are addressed by the strategies summarized below; full details,
including the explicit expressions of the quadrature weights, the
mapping functions, the terminating Kummer polynomial, and the
thresholds, are given in Section~5 of the SI.

\subsection{Catastrophic cancellation in the angular momentum coupling}
Angular-momentum coupling imposes exact cancellations whose depth
exceeds the working precision at two sites. The first is the
$l$-summation of the coupling table $J_{p_xp_yp_z}$: for same-centre
shell pairs with intermediate reduced exponents ($\mu\approx 5$--$25$,
typically the $d$-shell sets of second-row-and-heavier atoms), the
individual terms reach magnitudes of $\sim 10^{12}$ with alternating
signs, while their exact sum is of order $10^{-4}$. In IEEE binary64
($\epsilon\approx 10^{-16}$) each term already carries a rounding
error of order $10^{-4}$, so the accumulated error of the sum is of
order $1$--$100$, seven to nine orders of magnitude larger than the
quantity being computed. The second site is the $A$-coefficient cycle
and the quartet contraction, where the Hermite/Cartesian conversion,
the Gaussian prefactors, the products $A_xA_yA_zJ$, and the shell-pair
assembly involve per-pair cancellation depths of
$10^{12}$--$10^{15}$. In the spherically averaged intracule norm of
Kr$_2$ these cancellations, if evaluated in binary64, produced
relative errors growing with the internuclear separation, up to
$+24.7\%$ at $R=19.26$~bohr, oscillating in sign with $R$; lighter
systems (H$_2$, LiH, H$_2$O, Ne$_2$) remained accurate because they do
not exercise these channels.

The remedy follows from the elementary rule
$\epsilon_{\mathrm{working}}\times D < \mathrm{tolerance}$, where $D$
is the cancellation depth. For $D\lesssim 10^{16}$, binary64 leaves
$O(1)$ absolute noise, while binary128 ($\epsilon\approx 10^{-34}$)
leaves $\sim 10^{-19}$, orders of magnitude below any chemically
relevant tolerance. The whole cancelling chain must therefore be
elevated to IEEE binary128 (\texttt{real(16)}), not merely the
accumulator: widening only the accumulator is useless (the errors
enter through the terms), and a single binary64 boundary anywhere
inside the chain restores the binary64 floor regardless of how much of
the chain was widened around it. Quantities that are exact in binary64
(MO coefficients, RHF occupation numbers, integer density-matrix
elements) need no elevation. Since only some shell quartets
exercise these channels, the program decides quartet by quartet: each
quartet is first evaluated in binary64 together with a running bound
on its rounding error, and the whole chain is repeated in binary128
only for the quartets whose bound exceeds a relative tolerance
(input keyword \texttt{BINARY128 AUTO}, the default); binary128 can
also be imposed on every quartet (\texttt{BINARY128 ALWAYS}) or
switched off (\texttt{BINARY128 NEVER}). The three modes and the
error bound are described in Section~S5.5 of the SI. The affected sites, the validation
against \texttt{mpmath}, and the measured cost ($\le 2\times$, with a
Kr$_2$ STO-3G analytic-norm evaluation running in $\sim 11$~minutes on
8 threads) are documented in Section~5 of the SI.

\subsection{Numerical stability of the coupling functions}
The most delicate quantity is $B_l(\mu,R,T)$ of
Eq.~\eqref{besselscaled}. Since $i_l(2\mu RT)$ grows exponentially
while $e^{-\mu(R^2+T^2)}$ decays exponentially, evaluating the two
factors separately in binary64 leads to overflow or underflow whenever
$\mu(R^2+T^2)$ exceeds a few hundred. Rewriting the product as
\begin{equation}
B_l(\mu,R,T) = e^{-\mu(R-T)^2}\,i_l^{\mathrm{scaled}}(2\mu RT),
\label{eq:B_stable_practical}
\end{equation}
with $i_l^{\mathrm{scaled}}(z)=e^{-z}i_l(z)$, makes both factors
bounded and eliminates any overflow. Screening is then based on
$(R-T)^2$, not on $R^2+T^2$, which is much larger when
$R\approx T$, precisely where the contribution is significant, with
a threshold $\mu(R-T)^2>700$ whose robustness has been verified
empirically. The radial integral inside the $J$-build is evaluated as
a terminating Kummer polynomial of degree $m-1$, $m=(n-l)/2$, in
log-space binary128, with the prefactor as a closed-form
$\Gamma$-ratio and no $e^{-x}$ factor; this evaluation has been
validated against \texttt{mpmath} to $8.9\times 10^{-33}$ over the
full production range.

\subsection{Radial quadrature and validation}
Three strategies are used for the radial moments
$\int_0^\infty R^n I(R)\,dR$. The first is a standard quadrature
combined with a mapping $R=R(u)$ that transforms $[0,\infty)$ into a
finite range $u\in[a,b]$, using either the Euler--Maclaurin or the
Becke mapping. The second is the adaptive integrator \texttt{dqagi} of
\texttt{QUADPACK} \cite{piessens1983quadpack}, which increases the
number of $R$ values as needed until the estimated error falls below a
prescribed tolerance; this is particularly effective because $I(R)$
develops secondary maxima at values of $R$ corresponding to the
interatomic distances of the molecule, and a fixed grid may fail to
resolve them. The third is the exact analytical integration of
Sec.~\ref{sec:exact_radial}, which expresses the moments in closed form in terms of the
terminating Kummer polynomial. The three strategies are complementary
and their mutual agreement provides a stringent test of the
implementation.

The correctness of the elevated chain has been assessed against the
exact sum rules discussed in Sec.~\ref{sec:results}.
the analytic total norm for Kr$_2$ (STO-3G) equals
$N(N-1)=5112$ to $1.3\times 10^{-8}$ relative, uniformly flat across
the whole dissociation scan ($R=5.64$--$21.92$~bohr); all small
reference systems (H$_2$: $2.0$; LiH: $12.0$; H$_2$O STO-3G: $90.0$;
Ne$_2$: $380$) agree with their exact identities, and the
exchange-correlation channel reproduces $-N$ to the same relative
accuracy at every geometry. The remaining residual, $+6.64\times 10^{-5}$
absolute on 5112 at $R=8.06$~bohr (Table~\ref{kr2}), is not produced by the
integration. The molecular orbitals stored in the \texttt{.wfn} file, with eight
significant digits per coefficient, are orthonormal only to about $10^{-8}$, and
the electron and pair counts that they imply, $N'=\mathrm{tr}(\mathbf{DS})=72.00000047$,
$N'^2=5184.00006730$ and $-\tfrac12\mathrm{tr}(\mathbf{DSDS})=-72.00000093$, reproduce the
analytic values to all the digits given. The same holds for every system tested
(Section~S5.7 of the SI).

\section{Illustrative results}
\label{sec:results}

Perhaps the most straightforward way to assess the validity of the expression for $I(R)$ derived
in the preceding sections is to verify that it approximately satisfies
\begin{equation}
\label{eq:inticule_normmo}
\int_0^\infty I(R)\, dR \simeq N(N-1),
\end{equation}
where $N$ is the number of electrons of the molecule. Any deviation of the integral from the expected value $N(N-1)$
can only be ascribed to numerical inaccuracies inherent in its evaluation, since the sum over $l$ in Eq.~\eqref{angintj3} is rigorously limited to $l = p_x+p_y+p_z$ (Section~\ref{sec:methodradial}) and therefore introduces no truncation error. A less direct, albeit complementary,
way to validate the present formulation consists in verifying whether the total electron-electron repulsion energy
of the molecule, $V_{ee}$, computed by an independent method, is correctly recovered as
\begin{equation}
\label{eq:Vee}
V_{ee} \simeq \frac{1}{2} \int_0^\infty R^{-1} I(R)\, dR.
\end{equation}
Agreement between both members of the above equation would then lend further
credence to the overall theoretical framework, while systematic
discrepancies would point toward the need for a more refined numerical treatment
of the integrals or to the finite precision of the arithmetic employed in their evaluation.
The method can also be checked against the Coulomb (C) or Exchange-Correlation (XC) intracule density,
defined as
\begin{eqnarray}
\label{eq:rho2coulmo}
I^{\textrm{C}}(\bm{R})  & = & \sum_{ijkl} \Gamma^{\textrm{C}}_{ijkl}  I_{ijkl}(\bm{R}),\\
I^{\textrm{XC}}(\bm{R}) & = & \sum_{ijkl} \Gamma^{\textrm{XC}}_{ijkl} I_{ijkl}(\bm{R}),
\end{eqnarray}
where $\Gamma^{\textrm{C}}_{ijkl}$ and $\Gamma^{\textrm{XC}}_{ijkl}$ are the Coulomb and
Exchange-Correlation 2-RDM coefficients in the MO basis, defined as
$\Gamma^{\textrm{C}}_{ijkl}=\rho_{ij} \rho_{kl}$,
where $\rho_{ij}$ are the coefficients of the 1-RDM in the MO basis,
$\rho(\bm{r}_1;\bm{r}_1')=\sum_{i,j} \rho_{ij} \chi_i^\star(\bm{r}_1) \chi_j(\bm{r}_1')$, and
$\Gamma^{\textrm{XC}}_{ijkl}=\Gamma_{ijkl}-\Gamma^{\textrm{C}}_{ijkl}$.
Finally, $I^{\textrm{XC}}(\bm{R})$ can be divided in two terms, $I^{\textrm{XC}}(\bm{R})=I^{\textrm{XC,DF}}(\bm{R})+
I^{\textrm{Corr}}(\bm{R})$, with $I^{\textrm{XC,DF}}(\bm{R})$ (DF stands for Dirac-Fock) given by
\begin{eqnarray}
\label{eq:rho2df}
I^{\textrm{XC,DF}}(\bm{R}) & = & \sum_{ijkl} \Gamma^{\textrm{XC,DF}}_{ijkl} I_{ijkl}(\bm{R}),
\end{eqnarray}
where $\Gamma^{\textrm{XC,DF}}_{iljk}=-\frac{1}{2}\rho_{ij} \rho_{kl}$.

The integration of $I^{\textrm{C}}(R)$ and $I^{\textrm{XC}}(R)$ from 0 to $\infty$
should give the values $N^2$ and $-N$, respectively, one half the integrals of
$R^{-1} I^{\textrm{C}}(R)$ and $R^{-1} I^{\textrm{XC}}(R)$ must ideally give
$V_{\textrm{C}}$ and $V_{\textrm{XC}}$, the full molecular Coulomb and XC energies.

\subsection{Exactness of the integrated intracule density and sum rules}
Table~\ref{varias_molecules} collects the integrated intracule density \(\int_0^\infty I(R)\,dR\) for a diverse set of molecules (C\(_4\)H\(_4\), Ne, OH\(^-\), F, N, Li\(_2\), CO, LiH, F\(_3^-\), CH\(_4\)) computed at the RHF, ROHF, CAS and CCSD levels listed in the table. The results obtained via the analytical (confluent hypergeometric) radial integration and those obtained via the \texttt{QUADPACK} adaptive quadrature are compared with the exact sum rule \(N(N-1)\). For all systems, the analytical values reproduce the electron-pair count to at least seven significant digits (e.g., Ne: \(90.00000011\); F: \(72.00000033\); CO: \(182.00000359\); C\(_4\)H\(_4\): \(756.00000720\)). This level of agreement is particularly remarkable for the heavier systems, where the number of contributing quartets is large and the cancellation between positive and negative contributions is severe. Since both routes share the analytical angular average, their agreement validates the radial integration; the exactness of the angular part rests on the rigorous termination of the Bessel expansion at \(l = p_x + p_y + p_z\) (Sec.~\ref{sec:methodradial}) and on the comparison with Lebedev quadratures discussed below.

\begin{table}[htbp]
\centering
\caption{Integrated intracule density \(\int_0^{\infty} I(R)\,dR\) for a
representative set of molecules at different levels of theory, obtained via
\texttt{quadpack} and ``Analytic'' (confluent hypergeometric functions) radial
integration, both combined with analytic Bessel angular integration.}
\label{varias_molecules}
\begin{tabular}{llccc}
\hline
Molecule   & level &  \texttt{quadpack}  &    Analytic   & $N(N-1)$ \\
\hline
C$_4$H$_4$ & RHF/3-21G            & 756.00000723 & 756.00000720  & 756\\
Ne         & RHF/6-311G           &  90.00000011 &  90.00000011  &  90\\
OH$^-$     & RHF/TZV(2d,1p)       &  89.99999986 &  89.99999984  &  90\\
F          & ROHF/6-311G(d)       &  72.00000033 &  72.00000033  &  72\\
N          & ROHF/6-311G(d)       &  41.99999989 &  41.99999989  &  42\\
Li$_2$     & CAS(2,8)/6-311G(d)   &  30.00000003 &  30.00000003  &  30\\
CO         & CCSD/cc-pVDZ         & 182.00000359 & 182.00000359  & 182\\
LiH        & RHF/STO-6G           &  11.99999974 &  11.99999974  &  12\\
F$_3^-$    & RHF/6-311G(d)        & 756.00006066 & 756.00003114  & 756\\
CH$_4$     & RHF/6-311G(d,p)      &  89.99999978 &  89.99999978  &  90\\
\hline
\end{tabular}
\end{table}

\begin{figure}
\begin{center}
\includegraphics[width=0.99\columnwidth]{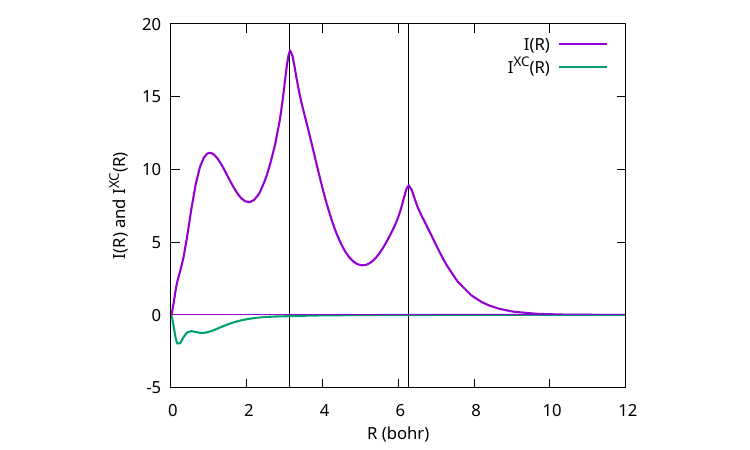}
\end{center}
\caption{Intracule density $I(R)$ and exchange--correlation intracule density
$I^{\mathrm{XC}}(R)$ of F$_3^-$ at the RHF/6-311+G** level,
obtained via \texttt{quadpack} radial and Bessel analytic angular integration.
The two vertical lines indicate the two different F--F internuclear distances in the molecule.}
\label{f3intracule}
\end{figure}

It is worth examining the small deviations observed in
Table~\ref{varias_molecules} in some detail. For most molecules, the
\texttt{QUADPACK} and analytical values agree to all the digits shown. The only
appreciable discrepancy appears for F\(_3^-\), where the \texttt{QUADPACK}
value (\(756.00006066\)) differs from the analytical one (\(756.00003114\)) in
the fifth decimal place, i.e., at the \(4 \times 10^{-8}\) relative level. This
is precisely the behavior expected for a system with several interatomic
distances, where the integrand \(I(R)\) develops multiple maxima that a fixed
or even an adaptive numerical grid may fail to resolve with uniform accuracy.
This is illustrated in Figure~\ref{f3intracule}, where the radial intracule
density of F\(_3^-\) exhibits three well-defined maxima at \(R \approx 1.03\),
\(3.14\), and \(6.28\) bohr, corresponding to an intra-atomic peak ($R\approx 1.03$ bohr), a bonded
F-F peak  ($R\approx 3.14$ bohr), and a terminal F$\cdots$F peak ($R\approx 6.28$ bohr), respectively, as expected for a linear triatomic anion. A
numerical quadrature that is not sufficiently fine in the regions around these
maxima will necessarily introduce a non-uniform error in the integrated
density, which is consistent with the small residual observed in
Table~\ref{varias_molecules}. The analytical route, by contrast, is free of
this quadrature noise by construction, and its small deviations from the
exact sum rules are of the size expected from the eight-digit molecular-orbital
coefficients of the input files, not from the integration itself.

\subsection{Performance of numerical quadratures: Euler--Maclaurin vs.\ \texttt{QUADPACK}}

\begin{figure}
\begin{center}
\includegraphics[width=0.99\columnwidth]{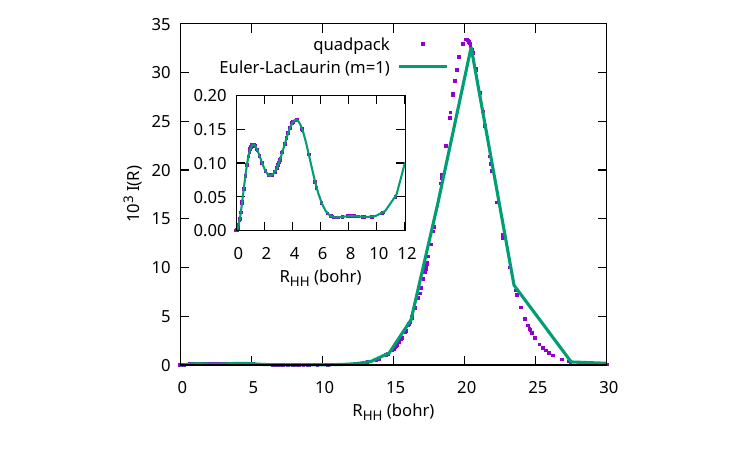}
\end{center}
\caption{\label{figure2}
Distribution of $R$ points in the Euler--Maclaurin with $m=1$ (EM1) and quadpack radial integrations of H$_2$ molecule ($R_{\textrm{HH}}=20.0$ bohr) at the full CI/aug-cc-pVTZ level.}
\end{figure}

The limitations of fixed-grid radial quadratures are illustrated in Table~\ref{tab:intracule} and Figure~\ref{figure2} for the H\(_2\) molecule at the full CI/aug-cc-pVTZ level (\(R_{\mathrm{HH}} = 20.0\) bohr). The Euler--Maclaurin mapping with \(m=1\) (EM1) yields a total integrated density of \(2.02416875\), a \(1.2\%\) deviation from the exact value of \(2.0\), and a \(V_{ee}\) of \(0.05225956\), which is about \(1\%\) above the exact value of \(0.05169870\). The Coulomb and exchange-correlation components show a similar pattern: \(V_{ee}^{\mathrm{Coul}}\) is \(0.47016293\) (vs.\ exact \(0.46960645\)) and \(V_{ee}^{\mathrm{XC}}\) is \(-0.41790337\) (vs.\ exact \(-0.41790774\)), while the Dirac--Fock component is \(-0.20917543\) (vs.\ exact \(-0.20917358\)). For \texttt{QUADPACK}, the only quantity reported in Table~\ref{tab:intracule} is the integrated density, \(2.00000011\), which agrees with the exact value \(2.00000000\) to \(1.1 \times 10^{-7}\).

\begin{table*}[htbp]
\centering
\caption{Comparison of the intracule density properties computed using the Euler--Maclaurin with $m=1$ (EM1) and \texttt{quadpack} radial integrations of H$_2$ molecule ($R_{\textrm{HH}}=20.0$ bohr) at the full CI/aug-cc-pVTZ level. The exact values were obtained with our domestic program \texttt{imolint}.}
\label{tab:intracule}
\begin{tabular}{lrcr}
\hline
Property & EM ($m=1$) & quadpack & Exact \\
\hline
	$V_{ee}$                             & 0.05225956    & -- & 0.05169870         \\
	$V_{ee}^{\textrm{Coul}}$             & 0.47016293    & -- & 0.46960645         \\
	$V_{ee}^{\textrm{XC}}$               & $-0.41790337$ & -- & $-0.41790774$      \\
	$V_{ee}^{\textrm{XC,DF}}$            & $-0.20917543$ & -- & $-0.20917358$      \\
$\int_0^{\infty} I(R)\,dR$                   & 2.02416875    & 2.00000011 & 2.00000000 \\
$\int_0^{\infty} I^{\textrm{Coul}}(R)\,dR$   & 4.02391969    & --         & 4.00000000 \\
$\int_0^{\infty} I^{\textrm{XC}}(R)\,dR$     & $-1.99975094$ & --         & $-2.00000000$ \\
\hline
\end{tabular}
\end{table*}

\begin{figure}
\begin{center}
\includegraphics[width=0.99\columnwidth]{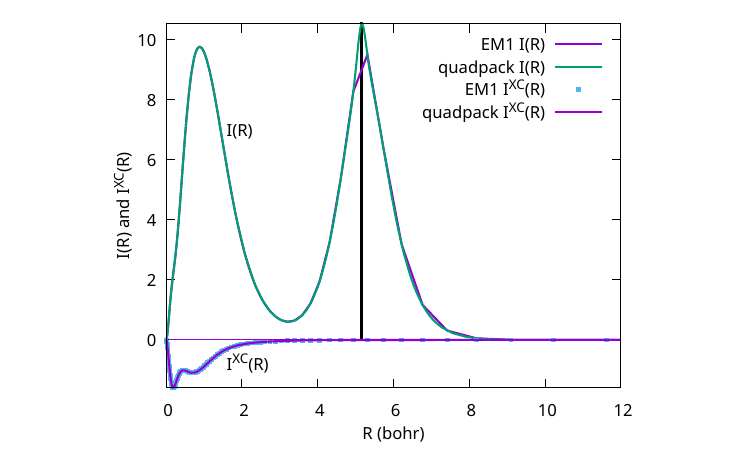}
\end{center}
\caption{\label{figure3}
Distribution of $R$ points in the Euler--Maclaurin with $m=1$ (EM1) and quadpack radial integrations of Ne$_2$ molecule ($R_{\textrm{NeNe}}=5.157$ bohr) at the CAS(16,12)/cc-pVTZ level of calculation. The vertical black line signals the internuclear distance of the molecule.}
\end{figure}

Figure~\ref{figure2} reveals the origin of the EM1 discrepancy: the EM1 grid
concentrates points at small \(R\) and fails to resolve the secondary maximum of
\(I(R)\) that develops at \(R \approx R_{\mathrm{HH}}\) upon dissociation.
\texttt{QUADPACK}, by contrast, adaptively places points precisely in this
region, as shown by the distribution of \(R\) points in Figure~\ref{figure2}.
The same conclusion is reached from Figure~\ref{figure3}, which shows the
distribution of \(R\) points for the Ne\(_2\) molecule at the
CAS(16,12)/cc-pVTZ level (\(R_{\mathrm{NeNe}} = 5.157\) bohr): the EM1 grid is
dense at short distances but sparse in the region around the internuclear
distance, where \(I(R)\) has a pronounced maximum, whereas \texttt{QUADPACK}
concentrates points exactly there. It should be noted, however, that this
deficiency of EM1 and similar fixed-grid schemes is much less serious for the
exchange-correlation intracule density \(I^{\mathrm{XC}}(R)\) and for
\(I^{\mathrm{XC}}(R)/R\). Unlike \(I(R)\), which exhibits pronounced maxima at
the interatomic distances of the molecule, \(I^{\mathrm{XC}}(R)\) is
short-ranged and does not develop such secondary maxima: it is concentrated at
small \(R\) and decays rapidly beyond a few bohr. Consequently, the low density
of EM1 points in the regions around the maxima of \(I(R)\) affects the
exchange-correlation components only weakly, and the corresponding errors
remain small even when the total norm and \(V_{ee}\) are poorly converged.

\begin{figure}
\begin{center}
\includegraphics[width=0.99\columnwidth]{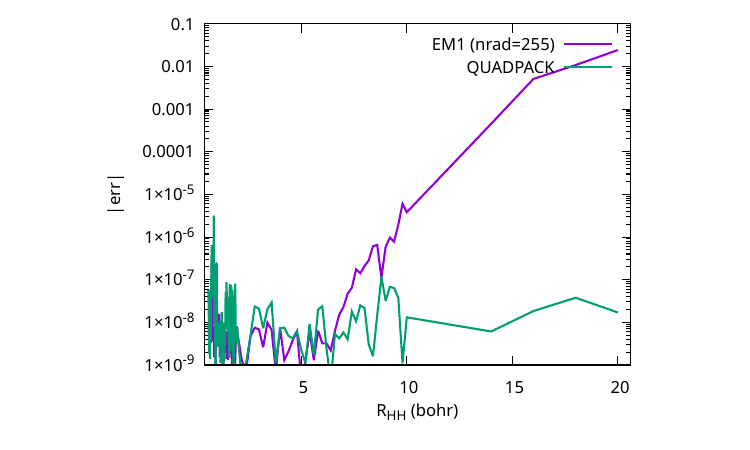}
\end{center}
\caption{\label{figure1}
Absolute value of the difference $\int_0^{\infty} I(R)\,dR-N(N-1)$ in the Euler--Maclaurin with $m=1$ (EM1) and \texttt{quadpack} radial integrations of H$_2$ molecule at the full CI/aug-cc-pVTZ level of calculation.}
\end{figure}

Figure~\ref{figure1} provides a quantitative view of the same effect. It displays the absolute value of the difference \(\int_0^\infty I(R)\,dR - N(N-1)\) for H\(_2\) as a function of the internuclear distance, for both EM1 and \texttt{QUADPACK}. The EM1 error grows rapidly as the bond is stretched, reaching values of order \(10^{-2}\) at \(R_{\mathrm{HH}} \approx 20\) bohr, while the \texttt{QUADPACK} error remains below \(10^{-6}\) over the entire scan. This is a direct consequence of the inability of a fixed radial grid to follow the secondary maximum of \(I(R)\) as it moves to larger \(R\) upon dissociation.

The convergence of the numerical quadratures is equally problematic for C\(_2\)H\(_6\) (RHF/STO-6G), as shown in Table~\ref{tab:C2H6_convergence}. Part (a) of the table shows that, with the analytical angular average, the EM1 norm agrees with the analytical value to within \(2\times10^{-6}\) from 150 radial points on, whereas \(V_{ee}\) converges much more slowly: with 255 points it is still \(5\times10^{-4}\) hartree below the exact value. The residual \(-1.8\times10^{-6}\) in \(\int I^{\mathrm{XC}}(R)\,dR\) is shared by the analytical route and is therefore not a quadrature error. Part (b) shows that the Lebedev angular convergence is worse: with 86 points the norm is \(328.79\) (a \(7.5\%\) error), with 434 points it is \(305.85\), and with 2030 points it is still \(305.98187\), i.e., about \(6 \times 10^{-5}\) (relative) away from the exact value of \(306\). Increasing the number of angular points to 5810 does not improve the result (\(305.98135\)), and the sequence of values is non-monotonic (230 points give \(281.11\), while 170 give \(313.82\)). The same erratic behavior is observed in \(V_{ee}\). Both the radial and the angular numerical quadratures therefore converge slowly and erratically for a molecule with several interatomic distances, which is precisely the motivation for the fully analytical angular integration developed in this work.

\begin{table}[htbp]
\centering
\renewcommand{\arraystretch}{0.95}
\caption{Convergence of intracule properties of the C$_2$H$_6$ molecule at the RHF/STO-6G level of calculation: (a) as a function of the number of radial points within the Euler--Maclaurin method with $m=1$; and (b) as a function of the number of Lebedev angular points with 100 radial points within the Euler--Maclaurin method with $m=1$. The ``Exact'' data were obtained using our in-house \texttt{imolint} program, and the ``Analytic'' values by using analytic radial and angular integrations as discussed in the text. Finally, the \texttt{quadpack} values required 255 radial points to achieve convergence and employed an analytic angular integration via Bessel functions. Wherever $\alpha_A$ is needed, it was taken as $1.455$ atomic units (Bragg radius of the carbon atom in diamond).}
\label{tab:C2H6_convergence}
\begin{tabular}{lcccc}
\toprule
\multicolumn{5}{l}{\textit{(a) Radial convergence: EM1 ($m=1$) + analytic angular (Bessel)}} \\
\hline
NR & $V_{\mathrm{ee}}$ & $V_{\mathrm{ee}}^{\mathrm{XC}}$ & $\int I(R)\,dR$ & $\int I^{\mathrm{XC}}(R)\,dR$ \\
\midrule
\hline
 50 &  68.93686775 & $-12.59801457$ & 305.99110418 & $-18.00000710$ \\
100 &  68.94838547 & $-12.60807329$ & 306.00009572 & $-18.00000216$ \\
150 &  68.95027298 & $-12.60997309$ & 306.00002848 & $-18.00000186$ \\
200 &  68.95094315 & $-12.61064294$ & 306.00003013 & $-18.00000180$ \\
255 &  68.95127596 & $-12.61097573$ & 306.00003023 & $-18.00000179$ \\
\midrule
\multicolumn{5}{l}{\textit{(b) Angular convergence: Lebedev + EM1 ($m=1$), 100 radial points}} \\
\hline
Lebedev & $V_{\mathrm{ee}}$ & $V_{\mathrm{ee}}^{\mathrm{XC}}$ & $\int_0^{\infty} I(R)\,dR$ & $\int_0^{\infty} I^{\mathrm{XC}}(R)\,dR$ \\
\midrule
\hline
  86 & 73.00109346 & $-12.60872820$ & 328.78890359 & $-18.00648756$ \\
 170 & 70.33254040 & $-12.60815778$ & 313.81912658 & $-18.00044451$ \\
 230 & 64.46782652 & $-12.60791511$ & 281.11216097 & $-17.99908338$ \\
 434 & 68.92240512 & $-12.60807412$ & 305.84518652 & $-18.00002222$ \\
 974 & 68.94571278 & $-12.60807327$ & 305.98710443 & $-18.00000196$ \\
2030 & 68.94507100 & $-12.60807318$ & 305.98187105 & $-18.00000128$ \\
3074 & 68.94501149 & $-12.60807317$ & 305.98140241 & $-18.00000119$ \\
4334 & 68.94500508 & $-12.60807317$ & 305.98135155 & $-18.00000119$ \\
5810 & 68.94500548 & $-12.60807317$ & 305.98135478 & $-18.00000119$ \\
\hline
\texttt{quadpack} & & & 306.00003036 & \\
Analytic & 68.95181082 & $-12.61151059$ & 306.00003022 & $-18.00000178$ \\
Exact    & 68.95181082 & $-12.61151059$ & 306.00000000 & $-18.00000000$ \\
\hline
\bottomrule
\end{tabular}
\end{table}

\subsection{The role of interatomic distances and the failure of fixed grids}

The limitations of fixed-grid quadratures are not restricted to molecules with
many atoms. Even a diatomic molecule such as CO exhibits a radial intracule
density with a rich structure, arising from the atomic shells and from the
bonding region, which a fixed radial grid may fail to resolve accurately. This
is illustrated in Table~\ref{tab:COconvergence}, which reports the relative
errors in the intracule properties of CO at the CCSD/cc-pVDZ level for
different angular and radial integration schemes. The table is organized so
that each method is described by three entries, corresponding to
\(\int_0^\infty I(R)\,dR\), \(V_{ee}\), and \(V_{ee}^{\mathrm{XC}}\),
respectively, and the columns labeled Leb86, Leb230, etc., indicate the number
of Lebedev angular points used in combination with the Bessel angular
integration (Bessel column) or with the numerical radial schemes (BMap+GL, EM1,
EM2, \texttt{QUADPACK}).

\begin{table*}[htbp]
\centering
\renewcommand{\arraystretch}{0.9}
\caption{Relative errors (in \%) in intracule properties of the CO molecule computed at the CCSD/cc-pVDZ level of theory using different angular and radial integration methods discussed in the text. BMap+GL: Becke mapping of $R$ onto the interval $u\in[-1,+1]$ combined with a Gauss--Legendre quadrature; EM1 and EM2: Euler--Maclaurin integration with $m=1$ and $m=2$, respectively; \texttt{quadpack}: radial integration performed with the \texttt{QUADPACK} package; Bessel column: angular integration carried out via Bessel functions. The analytic data denote the analytic radial and angular (Bessel) integrations discussed in the text. The reference exact values used to obtain the data were obtained with our in-house \texttt{imolint} program. The $n$ value in Leb$n$ denotes the number of angular points. For each integration method, the first, second, and third entries correspond to the $\int_0^{\infty} I(R)\,dR$, $V_{\mathrm{ee}}$, and $V_{\mathrm{ee}}^{\mathrm{XC}}$ values, respectively. In all cases except ``Analytic'' the number of radial points is 165. The value chosen for $\alpha_A$ in BMap+GL, EM1, and EM2 was of 2.46 atomic units.}
\label{tab:COconvergence}
\footnotesize\setlength{\tabcolsep}{3pt}
\begin{tabular}{lccccccc}
\hline
 & Bessel & Leb86 & Leb230 & Leb434 & Leb2030 & Leb4334 & Leb5810 \\
\hline
BMap+GL & $-0.544\times10^{-5}$ & 9.250 & $-9.060$ & $-0.036$ & $-0.004$ & $-0.004$ & $-0.004$ \\
 & $-0.490\times10^{-5}$ & 6.208 & $-6.082$ & $-0.024$ & $-0.003$ & $-0.003$ & $-0.003$ \\
 & $-0.220\times10^{-6}$ & 0.022 & $-0.003$ & $-0.220\times10^{-5}$ & $-0.294\times10^{-6}$ & $-0.294\times10^{-6}$ & $-0.294\times10^{-6}$ \\
\hline
EM1 & $0.203\times10^{-5}$ & 9.250 & $-9.070$ & $-0.036$ & $-0.004$ & $-0.004$ & $-0.004$ \\
 & $-0.103\times10^{-1}$ & 6.198 & $-6.093$ & $-0.034$ & $-0.013$ & $-0.013$ & $-0.013$ \\
 & $-0.480\times10^{-1}$ & $-0.026$ & $-0.051$ & $-0.489\times10^{-1}$ & $-0.480\times10^{-1}$ & $-0.480\times10^{-1}$ & $-0.480\times10^{-1}$ \\
\hline
EM2 & $0.304\times10^{-5}$ & 9.250 & $-9.07$ & $-0.036$ & $-0.004$ & $-0.004$ & $-0.004$ \\
 & $0.267\times10^{-5}$ & 6.209 & $-6.083$ & $-0.024$ & $-0.003$ & $-0.003$ & $-0.003$ \\
 & $0.735\times10^{-7}$ & 0.022 & $-0.003$ & $-0.184\times10^{-5}$ & $0.000\times10^{0}$ & 0.000 & $0.000\times10^{0}$ \\
\hline
\texttt{quadpack} & $0.197\times10^{-5}$ & 9.250 & $-9.070$ & $-0.036$ & $-0.004$ & $-0.004$ & $-0.004$ \\
 & $0.288\times10^{-6}$ & 6.208 & $-6.083$ & $-0.024$ & $-0.003$ & $-0.003$ & $-0.003$ \\
 & -- & -- & -- & -- & -- & -- & -- \\
\hline
Analytic & $0.197\times10^{-5}$ & & & & & & \\
 & $0.480\times10^{-7}$ & & & & & & \\
 & $-0.220\times10^{-6}$ & & & & & & \\
\hline
\end{tabular}
\end{table*}

Several conclusions can be drawn from Table~\ref{tab:COconvergence}, whose
entries are relative errors expressed in percent. First, with the analytical
Bessel angular integration (Bessel column) the error in the norm is of the
order of \(10^{-6}\%\) for every radial scheme. Second, when the angular
integration is performed numerically, the error is dominated by the angular
grid and is the same for all radial schemes: with 86 Lebedev points the
relative error in the norm is \(9.25\%\), it changes sign with 230 points
(\(-9.07\%\)), and it stalls at \(-0.004\%\) from 2030 points on. Third, once
the angular error is removed, the radial scheme decides the accuracy of the
energies. EM1 with 165 points leaves errors of \(10^{-2}\%\) in \(V_{ee}\) and
\(5\times10^{-2}\%\) in \(V_{ee}^{\mathrm{XC}}\), whereas BMap+GL, EM2 and
\texttt{QUADPACK} bring them down to \(10^{-6}\)--\(10^{-7}\%\), and the fully
analytical route gives \(5\times10^{-8}\%\) for \(V_{ee}\). The analytical angular
integration thus removes the dominant source of error, and what remains is
controlled by the radial quadrature, which must be chosen with care.

\begin{figure}
\begin{center}
\includegraphics[width=0.99\columnwidth]{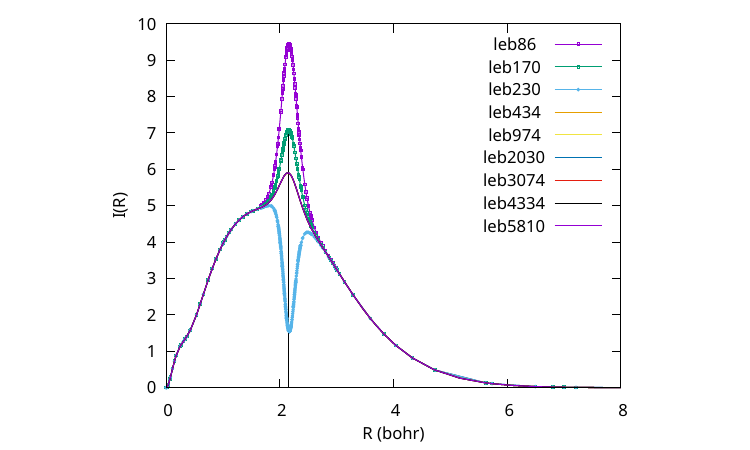}
\end{center}
\caption{\label{ccsd-co-ccpvdz}
Intracule density of the \ce{CO} molecule computed at the CCSD/cc-pVDZ level of theory. Angular integration was performed using numerical Lebedev quadratures (leb) with varying numbers of points, while the radial part was treated with \texttt{quadpack} integration.}
\end{figure}

Figure~\ref{ccsd-co-ccpvdz} illustrates the same point graphically for CO. The
intracule density develops multiple peaks corresponding to the C--O bond length
and the atomic shells, and the numerical Lebedev quadratures with insufficient
points (Leb86, Leb230) fail to resolve these features, producing large
errors at the main maximum, near the C--O bond length. The analytical Bessel angular
integration, by contrast, is free of this quadrature noise by construction, as
confirmed by the
relative errors collected in Table~\ref{tab:COconvergence}.

\subsection{Numerical stability in heavy systems: the Kr\(_2\) case}

The most stringent test of the numerical stability of the analytical formulation is provided by the Kr\(_2\) molecule at the RHF/STO-3G level. As discussed in Section~\ref{sec:practical}, the angular-momentum coupling in this system involves cancellations of depth \(10^{12}\)--\(10^{15}\), and a naive binary64 implementation produces relative errors in the norm of up to \(+24.7\%\) at \(R = 19.26\) bohr, oscillating in sign with \(R\). Lighter systems (H\(_2\), LiH, H\(_2\)O, Ne\(_2\)) remain accurate because they do not exercise these cancellation channels.

Table~\ref{kr2} shows the results obtained with the elevated binary128 chain described in Section~\ref{sec:practical}. The analytic radial integration (Exact $R$ integ) reproduces the exact norm \(N(N-1) = 5112\) to \(1.3 \times 10^{-8}\) relative, uniformly across the entire dissociation scan (\(R = 5.64\)--21.92 bohr), and the exchange-correlation component is exact at every geometry. The \texttt{QUADPACK} radial integration, combined with the analytic Bessel angular integration, yields \(5111.99956675\), in agreement with the exact value to \(8.5 \times 10^{-8}\) relative. The EM1 scheme with 100 radial points, by contrast, overestimates the norm by \(22\%\) (\(6237.94\)), and even with 1050 points the error remains at \(-0.1\%\) (\(5108.97\)). The exchange-correlation component is much less sensitive to the radial quadrature (its error is \(0.0\%\) even with 100 EM1 points), because \(I^{\mathrm{XC}}(R)\) is short-ranged and concentrated at small \(R\), where the EM1 grid is dense. The Coulomb component, by contrast, follows the same pattern as the total norm, with errors of \(21.7\%\) (EM1, 100 points) and \(-0.1\%\) (EM1, 1050 points).

\begin{table*}[htbp]
\centering
\caption{Relative error of intracule density properties for the Kr$_2$ molecule computed at the RHF/STO-3G level of theory using angular integration
via Bessel functions, together with three radial schemes: Euler--Maclaurin with
$m=1$ (EM1) using 100 and 1050 radial points, and the \texttt{quadpack} method
(which reached 1050 radial points). Exact $R$ integ denotes the analytic
radial integration discussed in the text.}
\label{kr2}
\begin{tabular}{|l|c|c|c|}
\hline
\hline
  & $\int_0^{\infty} I(R) dR$   &      $\int_0^{\infty} I^{XC}(R) dR$ &  $\int_0^{\infty} I^{Coul}(R) dR$ \\
\hline
Exact                &   5112                  &   -72                 &   5184   \\
EM1 (100)            &   6237.93519594 (22.0)  &   -72.00000647 (0.0)  &   6309.93520241 (21.7) \\
EM1 (1050)           &    5108.97320956 (-0.1) &   -72.00000094 (0.0)  &   5180.97321050 (-0.1) \\
Exact $R$ integ      &    5112.00006637 (0.0)  &   -72.00000093 (0.0)  &   5184.00006730 (0.0) \\
quadpack             &    5111.99956675 (0.0)  &    -- & -- \\
\hline
\end{tabular}
\end{table*}

\begin{figure}
\begin{center}
\includegraphics[width=0.99\columnwidth]{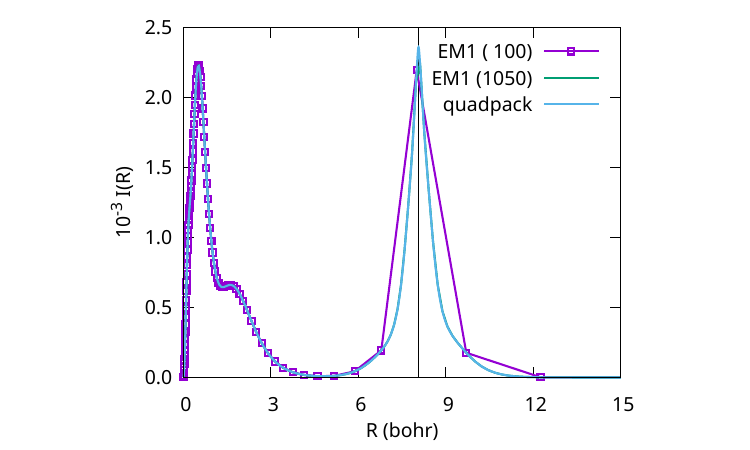}
\end{center}
\caption{\label{krfigure}
Intracule density of the Kr$_2$ molecule computed at the RHF/STO-3G level of theory. Angular integration was performed analytically via Bessel functions, while the radial part was treated with three different quadrature schemes: Euler--Maclaurin with $m=1$ (EM1) using 100 radial points, EM1 with 1050 radial points, and \texttt{quadpack} integration (which reached 1050 radial points). The vertical line marks the internuclear distance.}
\end{figure}

Figure~\ref{krfigure} shows the corresponding radial distributions for Kr\(_2\). The EM1 grid with 100 points fails to capture the tail of \(I(R)\), and the resulting curve deviates visibly from the \texttt{QUADPACK} results beyond \(R \approx 8\) bohr. With 1050 points, the EM1 curve is essentially superimposed on the \texttt{QUADPACK} curve over the whole range, confirming that the radial quadrature is the limiting factor and that the analytical angular integration is already converged. The vertical line in the figure marks the internuclear distance of the molecule (\(R_{\mathrm{KrKr}} \approx 7.8\) bohr at the STO-3G level), which coincides with the main maximum of \(I(R)\).

\subsection{Illustrative intracule densities: F\(_3^-\) and the physical picture}

Finally, Figure~\ref{f3intracule} shows the radial intracule density \(I(R)\) and the exchange-correlation intracule density \(I^{\mathrm{XC}}(R)\) for the F\(_3^-\) anion at the RHF/6-311+G** level, computed with \texttt{QUADPACK} radial and Bessel analytic angular integration. The total \(I(R)\) exhibits three well-defined maxima at \(R \approx 1.03\), \(3.14\), and \(6.28\) bohr, corresponding, respectively, to the maximum of the atomic F intracule density, to the F--F bond length, and to the second-neighbor F$\cdots$F distance. This is a direct manifestation of the physical transparency of the intracule representation: the positions of the maxima of \(I(R)\) map onto the interatomic distances of the molecule (excluding H atoms \cite{ylla-matito-ramoscordoba-2026}), and their relative heights reflect the number of electron pairs at each separation.

The exchange-correlation intracule \(I^{\mathrm{XC}}(R)\) is negative and short-ranged, with a minimum at \(R \approx 0.8\) bohr, reflecting the Fermi hole that suppresses the probability of finding two electrons with parallel spins at short distances. Beyond \(R \approx 4\) bohr, \(I^{\mathrm{XC}}(R)\) is essentially zero, while \(I(R)\) still has a sizable value, indicating that the long-range part of the intracule is entirely given by its Coulomb (uncorrelated) component. The qualitative behavior of \(I^{\mathrm{XC}}(R)\) is consistent with the sum rule \(\int_0^\infty I^{\mathrm{XC}}(R)\,dR = -N = -28\), and the agreement between the \texttt{QUADPACK} and analytical values in Table~\ref{varias_molecules} (\(756.00006066\) vs.\ \(756.00003114\)) shows that the adaptive radial integration is converged to $4\times10^{-8}$.

\subsection{Summary of the numerical validation}

Taken together, the results in Tables~\ref{varias_molecules}--\ref{kr2} and Figures~\ref{f3intracule}--\ref{krfigure} establish three conclusions. First, the analytical radial integration (Section~\ref{sec:exact_radial}) is exact in principle and reproduces the sum rules \(N(N-1)\), \(N^2\), and \(-N\) to at least seven significant digits for all systems tested, the residual \(1.3 \times 10^{-8}\) relative deviation in Kr\(_2\) being carried by the eight-digit molecular-orbital coefficients of the input wavefunction rather than by the method. Second, the Bessel analytic angular integration eliminates the quadrature error that affects numerical Lebedev grids, particularly for molecules with multiple interatomic distances, and its truncation at \(l = p_x + p_y + p_z\) is rigorous. Third, the numerical stability of the analytical route, achieved through the binary128 elevation of the cancelling chain and the scaled Bessel representation \(e^{-\mu(R-T)^2} i_l^{\mathrm{scaled}}(2\mu RT)\), is sufficient to treat systems as heavy as Kr\(_2\) without loss of accuracy. The \texttt{QUADPACK} adaptive quadrature remains a valuable independent reference, but its accuracy is limited by the same multiple-maxima structure that defeats fixed grids, and it cannot replace the analytical route when high precision is required.

\section{Conclusions}
\label{sec:conclusions}
We have presented a general analytical formulation of the radial intracule
density $I(R)$ for atomic and molecular systems described by Cartesian Gaussian
basis sets and arbitrary second-order density matrices. Starting from the
Hermite Gaussian expansion of the products of Cartesian Gaussian functions, we
have shown that the six-dimensional integral defining $I_{abcd}(\bm{R})$
factorizes into three one-dimensional integrals, and that the angular average
over the orientation of the interelectronic vector $\bm{R}$ can be carried out
analytically rather than numerically. The central difficulty of the molecular
case is the coupling factor $e^{2\mu\bm{R}\cdot\bm{T}}$, where
$\bm{T}=\bm{Q}_{cd}-\bm{P}_{ab}$ is the separation between the two Hermite
centres, which mixes the radial and angular variables and is absent in the
atomic case.

The analytical route to $I(R)$ developed in this work is based on an expansion
of the coupling exponential in modified spherical Bessel functions and real
spherical harmonics. This decouples the angular variables at the price of
introducing a sum over the angular momentum $l$, which we have shown to
truncate rigorously at $l=p_x+p_y+p_z$, where the $p$'s are the Cartesian
powers arising from the Hermite expansion. In addition to this analytical
approach, a purely numerical angular quadrature has been retained as an
independent reference against which the analytical results can be validated.

Special attention has been devoted to the numerical stability of the
analytical method. Two difficulties arise in the molecular case:
catastrophic cancellation in the angular-momentum coupling, where
terms of magnitude $\sim 10^{12}$ cancel to a result of order
$10^{-4}$, and overflow or underflow in the coupling functions,
where $i_l(2\mu RT)$ and $e^{-\mu(R^2+T^2)}$ vary exponentially in
opposite directions. The first is solved by elevating the whole
cancelling chain to IEEE binary128, following the rule
$\epsilon_{\mathrm{working}}\times D < \mathrm{tolerance}$, with the
radial integral evaluated as a terminating Kummer polynomial in
log-space binary128; the second, by rewriting the product as
$e^{-\mu(R-T)^2}i_l^{\mathrm{scaled}}(2\mu RT)$, which is bounded for
all values of $\mu$, $R$, and $T$.
When $\mu(R-T)^2$ exceeds a suitable threshold ($\simeq 700$), the corresponding
contribution is set directly to zero, and a logarithmic representation is
provided for the cases where the order of magnitude is still required.
The resulting implementation
reproduces, for Kr$_2$ across a full dissociation scan, the sum rules
implied by the input wavefunction to all the digits given; the
$1.3\times 10^{-8}$ relative deviation from the exact $N(N-1)$ is
carried by the eight-digit molecular-orbital coefficients of the
\texttt{.wfn} file.

In the atomic limit, where all Gaussian centres coincide and $\bm{T}=\bm{0}$,
the general formalism simplifies dramatically. Only the $l=0$ term of the
Bessel expansion survives, and the radial intracule density reduces to a finite
sum of terms of the form $C_i(\mu,\zeta)\,R^{n_i}e^{-\mu R^2}$, with only even
integer exponents $n_i$ ranging from 2 to $2+4l_{\max}$. This compact
representation allows the analytical evaluation of radial integrals, including
the normalization integral $\int_0^\infty I(R)\,dR$ and the electron-electron
repulsion integral $\int_0^\infty R^{-1}I(R)\,dR$, in terms of Gamma functions.
The same simplification applies to any quartet of Cartesian Gaussian functions
centred on a single atom of a polyatomic molecule, so that the atomic and
molecular formulations are connected smoothly and the atomic expression can be
recovered as a special case of the general one.

The correctness of the formulation has been assessed by comparing the values of
these radial integrals, computed with the analytical approach, and the
associated energetic components, namely the Coulomb, exchange-correlation, and
Dirac-Fock contributions, against reference values obtained from independent
sources. This comparison is a stringent test because the reference values
constitute an independent route to the same physical quantities. It is
important to emphasize, however, that in certain regimes the analytical
evaluation of these integrals is itself subject to numerical errors. This
happens, in particular, for basis sets of Cartesian Gaussian functions with
exponents of very large magnitude, and for very large values of the radial
variable $R$, where the analytical formulas involve products of large powers
and exponentials whose evaluation in finite precision may lose accuracy. In
such situations, the comparison between the two routes must be read as a
consistency check rather than as an absolute measure of the error of either
method, and small discrepancies may reflect the numerical limitations of the
analytical expressions rather than a defect of the present formulation.

With this caveat in mind, the zeroth moment of $I(R)$ reproduces the
electron-pair count $N(N-1)$, and the $(-1)$-moment reproduces the
electron-electron repulsion energy $V_{ee}$. The analogous sum rules for the
Coulomb and exchange-correlation intracule densities, namely $N^2$ and $-N$ for
the zeroth moments, and $V_{\mathrm{C}}$ and $V_{\mathrm{XC}}$ for the
$(-1)$-moments, provide additional checks that are sensitive to different
components of the second-order density matrix. The decomposition of the
exchange-correlation intracule into a Dirac-Fock term and a correlation term,
together with the corresponding decomposition of the second-order density
matrix, offers a further diagnostic tool for the analysis of electron
correlation in molecules. Since the Bessel series terminates exactly,
systematic deviations from these sum rules can only come from the radial
or angular quadratures, when these are numerical, or from the finite
precision of the arithmetic, and the comparison between the analytical and
numerical routes allows these sources of error to be disentangled.

The present formulation also complements earlier algorithmic developments in
this field. In particular, the approach of Cioslowski and Liu, based on
precomputed intermediates and two-stage integral screening, demonstrated that
the cost of evaluating intracule and extracule densities on large grids can be
reduced dramatically by reorganizing the calculation. Their algorithm remains
essentially numerical in the angular variable, whereas the formulation
presented here removes the angular quadrature altogether in the molecular case
by carrying out the angular average analytically. The two approaches are
therefore complementary: the algorithmic ideas of Cioslowski and Liu are well
suited to the rapid evaluation of the densities on very large grids, while the
analytical expressions derived here provide a route to the radial density that
avoids quadrature error and permits the analytical evaluation of radial
moments.

Several directions for future work can be envisaged. The formulation presented
here is general with respect to the second-order density matrix, and it can
therefore be applied without modification to any method that provides a
second-order density matrix, from Hartree-Fock to multiconfigurational and
coupled-cluster wavefunctions, as well as to the Coulomb, exchange, and
correlation components separately. The extracule density and the
momentum-space intracule follow a closely related algebraic structure and
will be considered in a separate publication. Finally,
the analytical expressions derived here are well suited to the evaluation of
higher radial moments and of the corresponding quantities in the framework of
explicitly correlated methods, where the interelectronic distance plays a
central role.

\begin{acknowledgement}
We acknowledge financial support from Grant PID2024-155569NB-I00, funded by \url{MICIU/AEI/10.13039/501100011033} and by the European Union (``ERDF A way of making Europe'').
Generative AI (Claude Opus 5.5, Anthropic) was used for editing the text and for checking the manuscript; the authors verified all results and take full responsibility for the content.
\end{acknowledgement}

\section*{Data and Software Availability}
The Fortran source code of the program used in this work, together with example input files and outputs, is available at \url{https://github.com/QTCOVI/intracule}.

\begin{suppinfo}
Evaluation of $I_{abcd}(R)$ for $s$-type functions; closed form of the atomic case; the angular integrals $G_{lm}$; derivation of the analytical radial moments; numerical precision and the binary128 chain; practical aspects of the implementation and thresholds (PDF).
\end{suppinfo}

\bibliography{intracule}

\end{document}